\documentclass{aastex631}

\pdfoutput=1
\usepackage{booktabs}
\usepackage{chemfig}
\usepackage{mathtools}

\usepackage{xcolor}

\received{May 27, 2021}
\revised{Mar 28, 2022}
\accepted{\today}

\submitjournal{ApJ}

\shorttitle{Comparing comet compositions with protosolar nebula models}
\shortauthors{Willacy et al.}

\graphicspath{{./}{figures/}}
\usepackage{graphicx}
\usepackage{tikz,pgfplots}
\pgfplotsset{compat=1.16}
\begin{document}

%%% This is necessary for drawing plot lines in the text
\scalebox{0}{
\begin{tikzpicture}
\dimendef\prevdepth=0
    \begin{axis}[hide axis]
        \addplot [
        color=black,
        solid,
        line width=0.9pt,
        forget plot
        ]
        (0,0);\label{hwplot1}
        \addplot [
        color=black,
        dashed,
        line width=1.2pt,
        forget plot
        ]
        (0,0);\label{hwplot2}
        \addplot [
        color=black,
        dashdotted,
        line width=1.2pt,
        forget plot
        ]
        (0,0);\label{hwplot3}
    \end{axis}
\end{tikzpicture}}

\title{Comets in context: Comparing comet compositions with protosolar nebula models} 

\correspondingauthor{Karen Willacy}
\author[0000-0001-6124-5974]{Karen Willacy}
\affil{Jet Propulsion Laboratory, California Institute of Technology, MS 169-506, 
4800 Oak Grove Drive
Pasadena, CA 91109, USA}

\author[0000-0001-8292-1943]{Neal Turner}
\affil{Jet Propulsion Laboratory, California Institute of Technology, MS 169-506, 
4800 Oak Grove Drive
Pasadena, CA 91109, USA}

\author[0000-0002-6391-4817]{Boncho Bonev}
\affil{Department of Physics, American University, Washington D.C., USA}

\author[0000-0003-0142-5265]{Erika Gibb}
\affil{Department of Physics and Astronomy, University of Missouri-St Louis, St Louis, MO, USA}

\author[0000-0002-8379-7304]{Neil Dello Russo}
\affil{Johns Hopkins Applied 
Physics Laboratory, 11100 Johns Hopkins Road, Laurel, MD 20723, USA}

\author[0000-0001-8843-7511]{Michael DiSanti}
\affil{Solar System Exploration Division, Planetary Science Laboratory Code 693, NASA/Goddard Space Flight Center, Greenbelt, MD, USA}
\affil{Goddard Center for Astrobiology, NASA/Goddard Space Flight Center, Greenbelt, MD, USA}

\author[0000-0002-8227-9564]{Ronald J. Vervack, Jr.}
 \affil{Johns Hopkins Applied Physics Laboratory, 11100 Johns Hopkins Road, Laurel, MD 20723, USA}

\author[0000-0002-6006-9574]{Nathan X. Roth}
\affiliation{Solar System Exploration Division, Astrochemistry Laboratory Code 691, NASA Goddard Space Flight Center, 8800 Greenbelt Rd, Greenbelt, MD 20771, USA}
\affiliation{Department of Physics, The Catholic University of America, 620 Michigan Ave., N.E. Washington, DC 20064, USA}

\begin{abstract}

Comets provide a valuable window into the chemical and physical
conditions at the time of their formation in the young solar
system. We seek insights into where and when these objects formed by comparing the range of abundances observed for nine molecules and their average values across a sample of 29 comets to the predicted midplane ice abundances from models of the protosolar nebula. Our fiducial model, where ices are inherited from the interstellar medium, can account
for the observed mixing ratio ranges of each molecule considered, but
no single location or time reproduces the abundances of all molecules
simultaneously.  This suggests that each comet consists of material
processed under a range of conditions. In contrast, a model where the initial composition of disk material is `reset', wiping out any previous chemical history, cannot account for the complete range of abundances observed in comets.

Using toy models that combine material processed under different thermal conditions we find that a combination of warm (CO-poor) and cold (CO-rich) material is required to account for both the average properties of the Jupiter-family  and Oort cloud comets, and the individual comets we consider. This could occur by the transport (either radial or vertical) of ice-coated dust grains  in the early solar system.

Comparison of the models to the average Jupiter-family and Oort cloud
comet compositions suggest the two families formed in overlapping
regions of the disk, in agreement with the findings of \cite{ahearn12} and with the predictions of the Nice model \citep{gomes05,tsiganis05}.

\end{abstract}

\keywords{Astrochemistry (75), Interstellar molecules (849), Comet origins (2203)}

\section{\label{sec:intro}Introduction}

Comets are among the best-preserved remnants of the early solar
system. They are the only class of primitive bodies to retain a rich
inventory of volatiles accreted in the cold outer regions of the
protosolar nebula.   Some molecular evolution may have occurred inside
comets since they formed, but despite this their primitive volatiles and easy accessibility for remote sensing investigations
make comets our best path to investigating the conditions in the icy
zone of the protosolar nebula during their epoch of formation. 

Comets likely
formed at diverse distances from the young Sun. The main scenarios for
their formation are (a) sequential agglomeration, where two-body collisions
build up successively larger masses
\citep{weidenschilling77,weidenschilling97,donn90,kataoka13},  (b) gravitational
collapse of a pebble cloud instigated by a mechanism such as the
streaming instability \citep[e.g.][]{yg05,simon16}, or (c) a combination of
the two \citep{davidsson16}.  Once formed, many nuclei were gravitationally
scattered by the young giant planets, sent either to the inner solar
system or to their present-day dynamical reservoirs: the Oort Cloud
\citep{oort50} and the scattered Kuiper Disk \citep{gladman05}.
Dynamical models of the young solar system based on the “Nice model” \citep{gomes05,tsiganis05}
suggest these two reservoirs were seeded by comets that formed in
overlapping regions, but in as yet unknown proportions. Today, various
processes 
gravitationally perturb individual comets from these
reservoirs, sending them to the inner solar system, where
sunlight warms the ices leading to the outgassing that enables their
compositions to be measured.

Measurements of cometary volatiles have been made with both space
probes and ground-based telescopes.
EPOXI and Rosetta provided in-situ measurements, and a
new space mission, Comet Interceptor, is being planned to focus
specifically on newly-discovered Oort cloud comets \citep{snodgrass19}. On the ground,
modern near-infrared spectrographs (iSHELL at IRTF, and NIRSPEC-2 at
Keck), as well as the sub-millimeter interferometer ALMA, are being
used for comet science. 
The primary motivation and overriding theme
for these extensive studies is the need for improved
understanding of comets as relics of the early solar system that
retain a rich inventory of volatiles from the cold regions of the
protosolar nebula \citep[][and references therein]{ahearn17}.

Like comet science, chemical modeling of protoplanetary disks
(PPDs) is
driven by the challenge of understanding the conditions for planetary
system formation.  (Protoplanetary disks around other stars are the external equivalent of the protosolar nebula). Modeling the disks' physical and chemical structure yields three distinct
 vertical regions: (1) a cold midplane, where ices freeze on to dust grains, and
where comet nuclei eventually form; (2) a warm molecular region, where
ices sublimate and are then processed via gas-phase reactions
involving radicals and ions, produced by the protostellar radiation
field; and (3) a hot ionized region containing predominantly atoms and
atomic ions \citep[Figure 1 in ][]{bergin07}.  Flows within the disk mix material 
between the layers, driving further molecular evolution.   
Historically, models have tended to be 
focused on interpreting the abundances of gas-phase molecules observed
in the
atmospheres of the PPDs around young stars, with less emphasis on
solid-phase volatiles in the protosolar nebula midplane and their links to comets. 
This trend has shifted in
recent years with new modeling projects tailored specifically to
predicting  the midplane abundances of volatiles
\citep{fa14,droz14,droz16,willacy15,kamp19,eistrup19}. Understanding
comets' connections to the early solar system requires combining simulations
treating the midplane ice inventory with ongoing efforts to disentangle
formative from evolutionary signatures in measured
abundances of comet volatiles \citep{bonev14,gibb17}. 

Here, we investigate the links between disk chemistry and comet
composition by modeling the evolution of the molecular abundances in
the ices found in the midplane of the solar nebula.  This work expands on previous studies by comparing the modeling
results  with observations of individual comets, as well as the  typical
composition of comets, considering the ensemble-averaged and range of abundances
across the observed objects for each
molecule. Diverse orbital and thus solar exposure histories mean that
post-formation processing may vary greatly even within each family. We
therefore compare the protosolar nebula models to the ensemble
properties first.  The disk modeling
approach is described in Section~\ref{sec:models} and the comet
observations in Section~\ref{sec:obs}. We discuss the results from
three models: the fiducial model  
(Section~\ref{sec:fiducial}), a model with a low cosmic ray
ionization rate (Section~\ref{sec:lowcr}), and   a model with atomic (`reset')
input abundances where it is assumed that any molecules formed in the molecular cloud are destroyed during the disk formation
(Section~\ref{sec:reset}).
In Section~\ref{sec:families} we consider whether the comparison
between models and observations suggest any systematic differences between 
Jupiter-family comets (JFCs) and Oort cloud comets (OCCs).
In Section~\ref{sec:1rad} we attempt to fit the average compositions
and those of individual comets in a
scenario where most of the comets consist of material from a single
place and time. Because this is not possible, in
Section~\ref{sec:2rad} we examine whether the ensemble and the
individual comets could have formed out of ices processed at different
locations in the protosolar nebula.  We discuss the findings'
ramifications in Section~\ref{sec:discussion} and summarize our
conclusions in Section~\ref{sec:concl}.

%%%%%%%%%%%%%%%%%%%%%%%%%%%%%%%%%%%%%%%%%%%%%%%%%%%%%%%%%%%%%%%%%%%%%%
\section{\label{sec:models}Disk Modeling Approach}

\subsection{\label{sec:diskmodel}Density and Temperature}

The disk's density and temperature structure is taken from a 1+1D
model
kindly provided by Paula d'Alessio and 
constructed using the methods set out in \citet{dalessio01}.  The total mass and its radial
distribution are similar to those of the minimum mass solar nebula.
The disk orbits a solar-mass star 2.6 R$_\odot$ in radius with
effective temperature 4000~K.  Mass accretes from disk to star at a
rate 2$\times$10$^{-8}$ M$_\odot$ yr$^{-1}$.  Well-mixed in the disk's
gas are dust particles following a power-law size distribution with
index $-3.5$ between 0.005~$\mu$m and 1~mm.  The temperature is
computed by balancing radiative cooling with heating by starlight and
by accretion power under the $\alpha$ viscosity prescription of
\cite{ss73}.  The temperatures of the gas and dust are assumed to be
equal inside the disk.  Resulting temperatures fall with distance,
$R$,  from
the star approximately as $R^{-0.5}$.  These parameters are
summarized in Table~\ref{tab:disk}.  The disk's density and
temperature structure remains fixed for the duration of the chemical
modeling and does not evolve with time.  We evolve the chemical
composition on the midplane between 1~au and 35~au.  The inner
boundary falls just inside the water snowline and the radial extent
covers the likely comet formation zone in the protosolar nebula.

\begin{deluxetable}{ll}
\tablecaption{\label{tab:disk}Star and disk parameters}
  \tablewidth{0pt}
    \tablehead{
    \colhead{Parameter} & \colhead{value}
    }
    \startdata
    $\alpha$ & 0.005\\
    M$_{*}$ & 1 M$_\odot$\\
    R$_*$   & 2.6 R$_\odot$\\
    T$_*$ & 4000 K \\
    \.{M} & 2 $\times$ 10$^{-8}$ M$_\odot$ yr$^{-1}$\\
    a$_{min}$ & 0.005 $\mu$m\\
    a$_{max}$ & 1 mm\\
    \enddata
\end{deluxetable}

\subsection{\label{sec:chem}Chemical Evolution}
Our chemical network is a subset of the 
UMIST database
\citep[RATE12;][]{rate12}, expanded to include gas-grain
interactions and grain surface reactions.  In addition, 
some reactions from the KIDA database \citep{kida} have been added to
extend the sulphur network and
the neutral-neutral chemistry in the gas-phase 
(Appendix~1).  The grain chemistry is calculated using the rate equation method. We use a `two-phase' model, in which the
composition of the gas and ice are followed, with the approach
described in \cite{cuppen17}. 
This modifies the rate equations to take into
account the number of ice monolayers on the grains and the competition
between reaction, diffusion and desorption.
We assume that the upper four layers of the ice mantle are chemically active and that the size of the barrier to surface diffusion is 0.35 $\times$ the binding energy, within the range suggested by the Monte Carlo simulations of \cite{kc14}.  H and H$_2$ are allowed to tunnel through any activation barriers.  

Ionization is driven by cosmic
rays, the cosmic-ray-induced photon field \citep{pt83}, and the decay of
radioactive nuclides. The effect of the stellar and interstellar
photon fields are ignored because of they don't penetrate to the
midplane.  For the cosmic ray ionization rate
($\zeta_{CR}$) we adopt the approach of 
\cite{semenov04}, where cosmic rays enter the disk from
both its top and bottom surfaces and are attenuated depending on the
surface density: 
\begin{equation}
\label{eq:cr}
  \zeta_{CR} = 0.5 \zeta_0 \left[exp(-\Sigma_1(z,R)/100) +
      exp(-\Sigma_2(z,R)/100)\right] \hbox{~~~s$^{-1}$}
\end{equation}
where $\Sigma_1$ and $\Sigma_2$ are the column densities (in
g cm$^{-2}$ to the top
and bottom of the disk respectively), and $\zeta_0$ is the 
interstellar cosmic ray ionization rate.  In our fiducial model we use
the standard interstellar value of \mbox{1.3 $\times$ 10$^{-17}$
s$^{-1}$}, whereas in a low-ionization version we use a value ten times
smaller.
The cosmic ray photodissociation rates from RATE12 are scaled to
account for the
change in $\zeta_{CR}$ compared to $\zeta_0$.
Ionization due to the decay of radioactive nuclides
 such as $^{26}$Al  has a rate given by  
 \begin{equation}
 \zeta_{Al} = 6.5 \times 10^{-19} exp(-0.693 t/0.73) \hbox{~~~s$^{-1}$} 
 \end{equation}
 where $t$ is the elapsed time in millions of years.  The exponential
 factor accounts for the decreasing abundance of $^{26}$Al as it
 decays with  half-life of 0.73~Myr \citep{cr09}.

\begin{deluxetable}{lcllc}
\tablecaption{\label{tab:be}Binding energies of important species.
  Values are taken from Table 3 in \cite{penteado17}.} 
\tablewidth{0pt} 
\tablehead{
\colhead{Species} & \colhead{Binding energy} & \colhead{  } &
\colhead{Species} & \colhead{Binding energy} 
} 
\decimals
\startdata
H                & 650 & & H$_2$        & 500 \\
CH$_4$      & 1250 & & CO             & 1100 \\
H$_2$CO    & 3260 & & CH$_3$OH  & 3820\\
CO$_2$       & 2267 & & H$_2$O       & 4800\\
N$_2$         & 990 & & NH$_3$       & 2715\\ 
OH              & 3210 & & O                & 1660 \\
O$_2$        & 898  & & C$_2$H$_2$   & 2090 \\
C$_2$H$_6$   & 2183 & & HCN          & 1583 \\
HNC          & 1510 & & OCS          & 2325 \\
CH            & 590 \\
\enddata
\end{deluxetable}

Reaction and diffusion rates on the grains depend on the ice species'
binding energies on the surface.  We use the energies in
Table~\ref{tab:be} from \cite{penteado17}.  Following
\citet{cuppen17}, only the ice's uppermost four monolayers are
available to desorb.  We assume molecules are returned to the gas by
cosmic ray heating of the grains and by thermal desorption.  In the
midplane, the ices are shielded from both interstellar and stellar
photons, but cosmic rays still penetrate, and their cascade yields
energetic photons that can photodesorb the ices and photodissociate
the ice molecules.  We assume the latter occurs at the same rate as
gas-phase photodissociation \citep{rh01}.

\subsection{\label{sec:initial}Initial Abundances}

The initial abundances of all chemical species in the fiducial disk
model come from a molecular cloud chemistry model.  We treat a
location in the cloud's interior where the temperature is 10~K,
the total hydrogen density  is 2$\times$10$^4$ cm$^{-3}$ and the visual
extinction is 10~magnitudes. 
We begin the cloud model with all elements in their atomic form except
carbon, which is present as C$^+$, and hydrogen, which is 1\%
atomic with the rest molecular (Table~\ref{tab:init}). The assumed C/O
ratio is 0.54, roughly solar.  We then evolve
the composition for 1~Myr.  Table~\ref{tab:ice} shows the resulting
ice abundances alongside the ices observed towards background stars
and in low-mass young stars.  Overall there is reasonable agreement
with both sets of observational data.

It is possible that energetic processing during disk formation could destroy any molecules that formed in the parent molecular cloud. In this case the disk chemistry would
start from ions and atoms rather than molecules.  The input
composition would be `reset', wiping out any record of the molecular
cloud chemistry.  To explore the implications for the results, we
compute an additional disk model, starting from such atomic
abundances (see Section~\ref{sec:reset}).  For this model the initial conditions are the same as for the cloud
model in Table~\ref{tab:init}.   In reality, a combination of the two scenarios is possible, with some molecules being destroyed and others surviving intact, e.g. \cite{lunine91,nh94, visser09}.  This possibility is not considered here. 

  \begin{deluxetable}{ll}
  \tablecaption{\label{tab:init}Initial abundances for the molecular
    cloud model given as a fractional abundance relative to total
    hydrogen (=n(H) + 2n(H$_2$)).  The C/O ratio is 0.54.} 
 \tablehead{
 \colhead{Species} & \colhead{Fractional}\\
\colhead{} & \colhead{abundance}
 }
 \startdata
 H & 1 (-2) \\
 H$_2$ & 4.95 (-1)\\
 He & 1.4 (-1) \\
 C$^+$ & 1.3 (-4)\\
 O & 2.4 (-4) \\
 N & 2.14 (-5) \\
 S$^+$ & 1.66 (-5)\\
 Si$^+$ & 8.0 (-9) 
 \enddata
 \end{deluxetable}
 
  \begin{deluxetable}{llllcc}
  \tablecaption{\label{tab:ice}Ice abundances calculated in molecular
    cloud model after 1~Myr.
    Also shown are the range of observed ice
    abundances towards background stars and low mass
    young stellar objects \citep[from Table 2 in
    ][]{boogert15}. Abundances are given as a percentage relative to
     water ice. \nodata indicates that no observational
    data are available. } 
  \tablewidth{0pt}
  \tablehead{
  \colhead{Molecule} & \colhead{Cloud model} & \colhead{Background stars} & \colhead{Low Mass YSO}
  }
  \decimals
  \startdata
  H$_2$O & 100 & 100 & 100 \\
  CO        & 40.5 & 9 -- 67  & $<$3 - 85\\
  CO$_2$ & 4.7 &  14 -- 43  & 12 - 50 \\
  CH$_4$    & 13.3 & $<$ 3 & 1 - 11\\
  CH$_3$OH  & 7.2 & $<$ 1 -- 12 & $<$1 - 25\\
  H$_2$CO   & 10.1 & \nodata & $\sim$ 6\\
  OCS       & 0.02 & $<$ 0.02 & $<$ 1.6\\
  NH$_3$    & 4.7  & $<$ 9 & 3 - 10\\ 
  HCN       & 3.2 & \nodata & \nodata\\
  C$_2$H$_2$ & 0.02 & \nodata & \nodata \\
  C$_2$H$_6$ & 1.8 (-3)  &  \nodata & \nodata \\
  \enddata
  \end{deluxetable}

\section{\label{sec:obs}Comet Sample}

Present-day cometary volatiles' composition could be affected by various
post-formation processes, short- or long-term.  Short-term processes
include diurnal and seasonal variations that are highly variable from
comet to comet.  Long-term processes are expected to depend strongly
on a comet's dynamical history.  However, there is compelling evidence
that the volatiles retain cosmogonic signatures.  First, the nucleus
of comet 67P/Churyumov-Gerasimenko shows near-solar abundances of
oxygen and carbon from in-situ measurements \citep{rubin19}.  Second,
the two fragments of the Jupiter-family comet 73P/SW3 have similar
compositions, in contrast to the diversity seen in the comet
population as a whole \citep{dr07}.  Similarly, the Oort cloud comets
Tabur (C/1996 Q1) and Liller (C/1988 A1), thought from their orbits to
be fragments of a single parent body, have similar compositions
\citep{turner99}.  Such fragments would be unlikely to share
compositions if the original nucleus suffered depth-dependent
processing during its numerous passages around the Sun.  Furthermore,
a study of product species based on optical measurements of 85~comets
shows no correlation between carbon-chain depletion and dynamical age,
suggesting the differences in carbon-chain chemistry among at least
some of the comets are natal \citep{ahearn95}.

Because the degree of post-formation processing is unknown and will be
different for each comet, we use an ensemble average of a sample of
comets as the first point of comparison for our protosolar nebula model.
The sample includes nine Jupiter-family comets (JFCs) and twenty Oort cloud
comets (OCCs) (Appendix~2).   The observations are
  taken from \cite{dr16} (hereafter DR16).
  
The species included in our study are: HCN, NH$_3$, H$_2$CO, CH$_3$OH,
C$_2$H$_2$, C$_2$H$_6$, CH$_4$, CO and H$_2$O (DR16).  We also include OCS whose abundances in several of our comets were recently reported by \cite{saki20}.  This selection of species is motivated, first, by their belonging to different chemical groups (symmetric hydrocarbons, oxygen-carbon compounds, nitrogen-bearing species), as optimal to connect with astrochemical models.  OCS is included because it
  forms a link between the carbon and oxygen chemistry and that of
  sulphur,
  and it is one of the few sulphur molecules to
  have been observed in a sufficient number of comets to allow an
  average to be determined.  
Second, the measurements of their relative abundances are based on (1) simultaneous observations of trace volatiles and H$_2$O, and (2) analysis with the same technique.  These  two conditions together result in the most reliable relative abundances as needed for comparison with disk models.  The list of species does not include CO$_2$, a major volatile, because it is commonly measured by different techniques and at different times.  Interpreting the CO$_2$ abundances can be a subject of a separate dedicated study.

We consider the mixing ratios averaged over the complete sample of 29
comets, as well as averages for the JFC and OCC
sub-samples individually (Appendix~2).   Full details
  of the observations and the calculation of the average compositions
  are given in DR16, but to summarize,  the averages used
are an unweighted mean of the mixing ratios with respect to water of a given species in each
comet.   Some observations (those in parentheses in Appendix~2) were excluded from the
calculation of the average composition for various reasons,
e.g. location of comet when measurements were taken, poor constraints
on the observations (see DR16 for details).

While this paper was in review a new set of measurements was published by Lippi et al. (2021) (hereafter L21), including 20 of the comets in the DR16 survey. Although in depth comparison between the two surveys is outside the scope of this paper, in Appendix~3 we carefully assess how the new L21 \citep[together with their earlier paper][; hereafter L20]{lippi20} results affect our analysis. None of the conclusions in this work, as presented in Section~\ref{sec:concl}, are changed by the introduction of the new survey.

Comparing the two
families, there are individual JFCs and OCCs with very similar abundances of volatiles.  The OCcs show wider ranges in the molecular mixing ratios relative to water, as well as higher average abundances.  Every species' mixing ratio spans a range of at least one order of magnitude. However, we emphasize JFCs are under-represented class in compositional studies of all parent species. Currently, they seem more depleted in the hyper-volatiles CO and CH$_4$ than OCCs, but very few measurements for these species have been feasible in JFCs, due to their weaker intrinsic brightness and Doppler-shift limitations  \citep{disanti17}.

\begin{deluxetable}{lccccccccc}
\tablecaption{\label{tab:av}Average observed molecular abundances as a percentage
  relative to water taken from Table 4 in DR16. Data for OCS
  comes from \cite{saki20}. The table shows the range of values measured
  (from lower limit to upper limit) and the average for the class of
  comets (Oort cloud and Jupiter-family comets). Letters
  indicate to which comet a particular observation refers.  Note that
  these ranges are based on relatively few measurements and so are
  unlikely to represent the true range in comets.  This is especially
  true where we have very few measurements (e.g. of CH$_4$ and OCS in
  JFCs).}  
\tablewidth{0pt}
\tablehead{
\colhead{molecule} & \multicolumn{3}{c}{OCC} & \multicolumn{3}{c}{JFC} & \colhead{All Comets}\\
\cmidrule(lr){2-4}\cmidrule(lr){5-7}\cmidrule(lr){8-8}
\colhead{} & \colhead{Lower limit} & \colhead{Average} & \colhead{Upper limit} & \colhead{Lower limit} & \colhead{Average} & \colhead{Upper limit} & \colhead{Average} 
}
\decimals
\startdata
CH$_3$OH & $<$ 0.2\tablenotemark{a} & 2.21 & 3.72\tablenotemark{b} & 0.54\tablenotemark{c} & 1.73 & 3.48\tablenotemark{d} &  2.06  \\ 
HCN & 0.07\tablenotemark{e}& 0.22 & 0.50\tablenotemark{f} & 0.03\tablenotemark{g} & 0.17 & 0.29\tablenotemark{c} & 0.21 \\
NH$_3$ & 0.10\tablenotemark{h} & 0.91 & 3.63\tablenotemark{i} & $<$ 0.09\tablenotemark{g} & 0.59 & 0.90\tablenotemark{j} & 0.80 \\
H$_2$CO & $<$ 0.04\tablenotemark{e} & 0.33 & 1.10\tablenotemark{j} & 0.13\tablenotemark{l} & 0.26 & 0.84\tablenotemark{j} & 0.31 \\
C$_2$H$_2$ & 0.04\tablenotemark{e} & 0.16 & 0.45\tablenotemark{l} & 0.03\tablenotemark{g} & 0.07 & 0.15\tablenotemark{m} & 0.13 \\
C$_2$H$_6$ & 0.26\tablenotemark{e} & 0.63 & 1.97\tablenotemark{f} & 0.12\tablenotemark{n} & 0.34 & 0.75\tablenotemark{k} & 0.55\\
CH$_4$ & 0.15\tablenotemark{a} & 0.88 & 1.57\tablenotemark{f} & $<$ 0.25\tablenotemark{o} & 0.31 & 0.54\tablenotemark{i} & 0.78 \\
CO & 0.4\tablenotemark{e} & 6.1 & 26.2\tablenotemark{p} & 0.3\tablenotemark{k} & 1.6 & 4.3\tablenotemark{i} & 5.20 \\
OCS & 0.04\tablenotemark{q} & 0.31 & 0.41\tablenotemark{r} & 0.06\tablenotemark{d} & 0.095 & 0.12\tablenotemark{n} & 0.17 \\
\enddata
\tablerefs{$^a$C/1999 S4, $^b$C/2007 N3, $^{c}$SW3-B, $^{d}$2P/Encke,
  $^e$8P/Tuttle, $^f$C/2007 W1, $^g$6P/d'Arrest,
   $^h$C/2013 R1,  $^i$C/2012 S1, $^j$9P/Tempel 1,
 $^{k}$103P/Hartley 2, $^l$C/2006 P1, $^{m}$81P/Wild, $^{n}$21P/G-Z, $^{o}$73P/SW3-C,
  $^{p}$C/1995 O1, $^q$C/2002 T7, $^r$Hale-Bopp}
\end{deluxetable}

\section{\label{sec:fiducial}Fiducial Model Results}  

In this section we present results from the fiducial protosolar nebula
model.  Sections~\ref{sec:lowcr} and \ref{sec:reset} respectively
cover a version where the cosmic ray ionization
rate is reduced by one order of magnitude and a version where the cosmic
ray rate is normal but all molecules except H$_2$ are initially
dissociated, with abundances given in Table~\ref{tab:init} (`reset' abundances).
Table~\ref{tab:models} summarizes the different models. 
The results are compared to
the average values and ranges of the mixing ratios from the complete
sample of JFCs and OCCs. 

\begin{deluxetable}{lll}
  \tablecaption{\label{tab:models}Summary of parameters used in the modeling.  The `inheritance' input abundances indicate that the initial abundances in the disk model are taken from the molecular cloud model described in Section~\ref{sec:initial}.} 
\tablewidth{0pt}
\tablehead{
\colhead{Model} & \colhead{Input abundances} & \colhead{$\zeta$}
}
\startdata
fiducial model & `inheritance' & $\zeta_0$\\
low ionization  & `inheritance' & 0.1$\zeta_0$\\
atomic & `reset'  & $\zeta_0$
\enddata
\end{deluxetable}

\begin{figure}
\centering
    \includegraphics[trim=275 20 20 20, clip, width=1.1\linewidth]{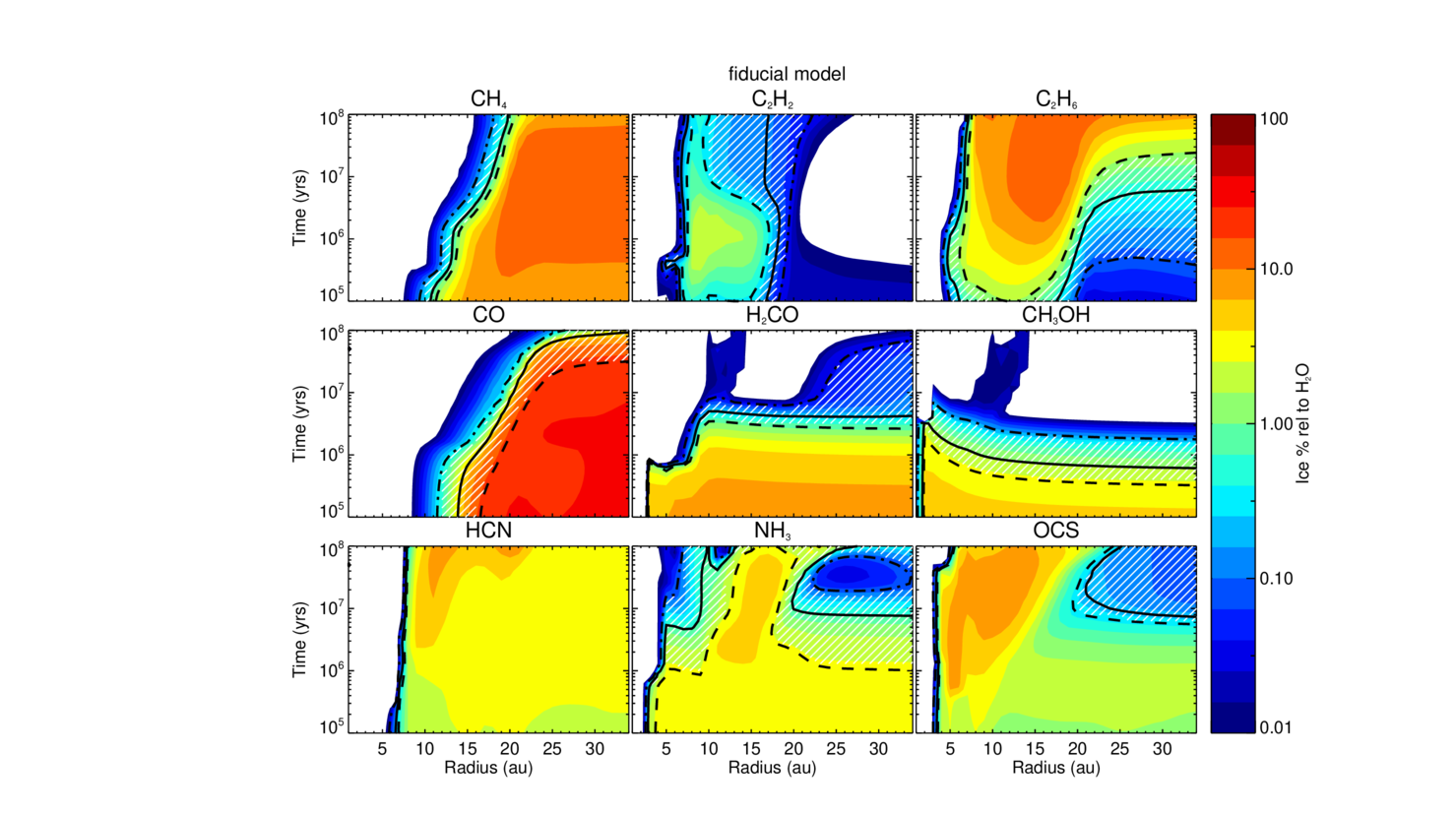}
    \caption{Ice abundances in the fiducial disk model compared with
      the averages and ranges for the observed comet values in Table~\ref{tab:av}.
      Color shading shows the modeled molecular ice abundances
      relative to H$_2$O.  The white hatched area indicates the range
      among the observed comets.  The upper end of the range is
      indicated with a black dashed \ref{hwplot2} line, the lower end
      by a black dot-dashed \ref{hwplot3} line.  The average observed
      composition is the solid black line.  Average observed values
      are from DR16, except OCS which comes from \cite{saki20}}.
    \label{fig:ratio_all}
\end{figure}

Figure~\ref{fig:ratio_all} shows the distributions of the ices as
predicted by the fiducial model,  overlaid with the corresponding abundances 
observed in the comets.   All are shown as  percentages of the 
water ice.  All nine molecules under consideration can be accounted
for by some combination of time and location in the model disk as follows.

\subsection{CO, H$_2$CO and CH$_3$OH}
The  CO/H$_2$O mixing ratio in the fiducial disk ranges from 0 -- 41\% 
depending on time and location.  The model matches the entire range of the
comet observations in a region close to the CO snowline.
The highest value of CO/H$_2$O observed in our comet sample is in Hale-Bopp.  A
weighted average value of 26.2\% was determined by
DR16 and this can be matched by the models around 17~au. The
highest individual measurement in this comet is 
41\% $\pm$ 13, similar to the molecular cloud model abundance and matched by
the disk model at $R$ $>$ 20 au.

The CO ice abundance does not increase in the disk model and the
molecular cloud value is maintained until this molecule is destroyed
either by reactions or desorption.
Thermal
desorption produces the 
CO snowline between 12 and 15~au, whereas the decrease at larger radii
after 1~Myr is due to cosmic ray desorption, dissociation by cosmic
ray-induced photons, and the  reactions:  
\begin{eqnarray}
    \hbox{CO:ice} + \hbox{NH:ice} & \longrightarrow & \hbox{HNCO:ice} \label{eq:hnco}\\
    \hbox{CO:ice} + \hbox{S:ice} & \longrightarrow & \hbox{OCS:ice}\label{eq:ocs}\\
    \hbox{CO:ice} + \hbox{O:ice} & \longrightarrow & \hbox{CO$_2$:ice}\label{eq:co2}
\end{eqnarray}

\begin{figure}
 \centering
  \includegraphics[width=0.7\linewidth]{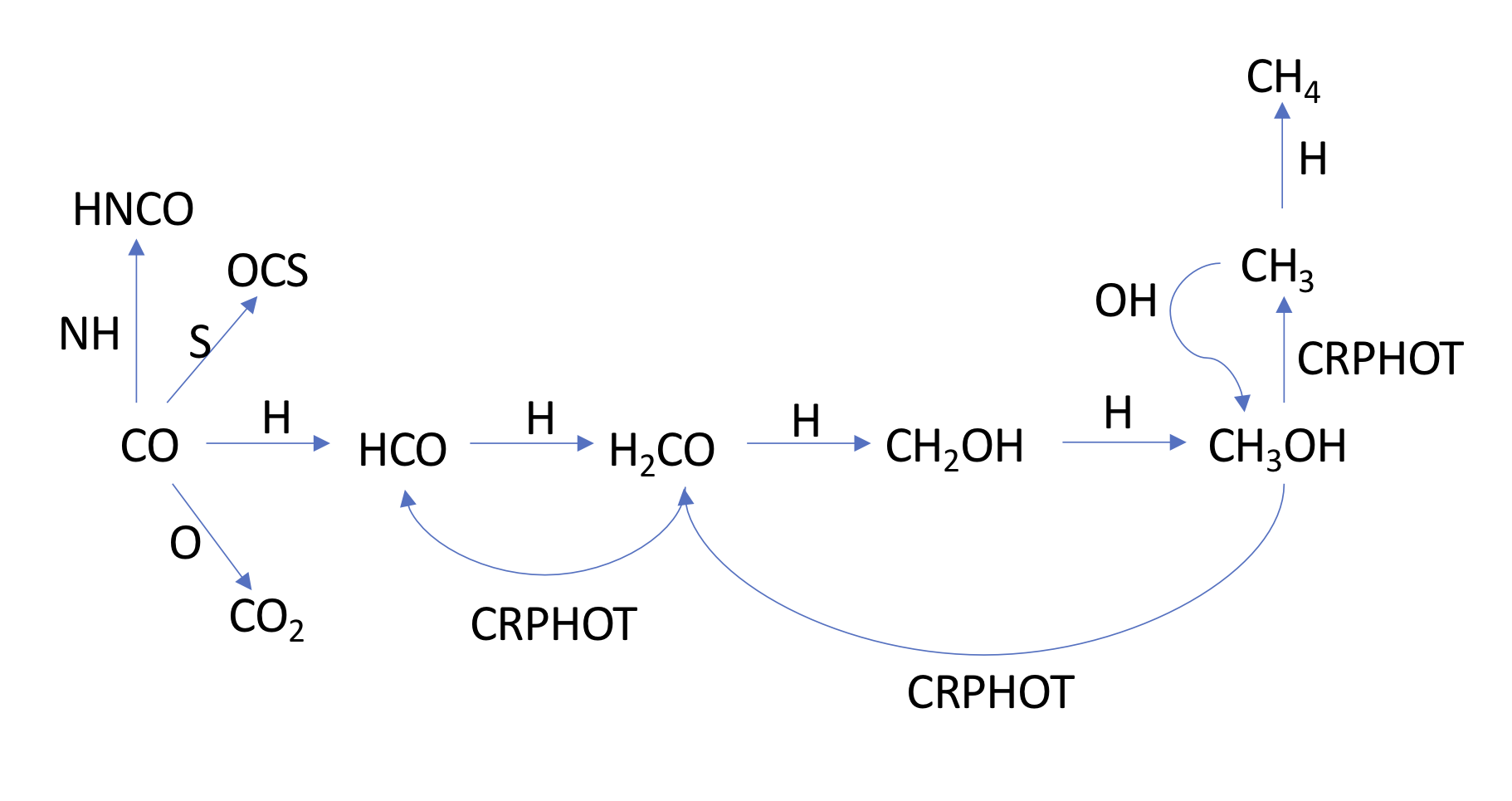}
 \caption{\label{fig:cochem}The grain chemistry of CO, H$_2$CO and
    CH$_3$OH.  All species shown are ice molecules. H$_2$CO and
    CH$_3$OH must be reduced from their molecular cloud abundances to
    fit the comet observations and this is achieved by processing of
    the ice by cosmic ray photons.}
\end{figure}

In the molecular cloud the mixing ratios of CH$_3$OH and H$_2$CO are
higher than observed in comets.  A good fit with the
observations is only found after these molecules' abundances are reduced by
processing in the disk. 
 Their destruction occurs mainly by cosmic ray
photons (Figure~\ref{fig:cochem}).  Although some recycling between
H$_2$CO and CH$_3$OH occurs, there is a gradual loss of these
molecules through
\begin{equation}
  \hbox{CH$_3$OH} \xrightarrow[]{\text{CRPHOT}} \hbox{CH$_3$} +
  \hbox{OH}
\end{equation}
CH$_3$ then goes on to form CH$_4$.
The observed range in CH$_3$OH/H$_2$O is matched by the model between
0.3 -- 3~Myr.  For H$_2$CO the fit is over a narrow range of times
from 2 -- 5~Myr at $R$ $<$ 20~au, and between 2 -- 60 Myr at $R$ $>$ 20~au.

Another way to reduce the abundance of H$_2$CO in particular might be
sequestration in a less volatile form, in which case the observed
mixing ratios would not reflect the total molecular abundance in the
comet. This possibility is discussed further in Section~\ref{sec:1rad}.

\subsection{CH$_4$}
Methane ice forms in the molecular cloud by hydrogenation of carbon
atoms and ions adsorbed onto the grains, and its abundance does not
change appreciably with time or with radius over much of the
disk. Only around its snowline can the observed cometary mixing
ratios be matched.

\subsection{C$_2$H$_2$ and C$_2$H$_6$}
Both C$_2$H$_2$ and C$_2$H$_6$ have low  abundances in the molecular
cloud, where C$_2$H$_2$/H$_2$O = 0.02 \% and the C$_2$H$_6$
ratio is a factor of 11 lower. The comet abundances range from 0.04 to
0.4 \% for C$_2$H$_2$/H$_2$O, and from 0.26 to 2 \% for C$_2$H$_6$.
Therefore to fit the observations, these two molecules must form in the
disk.

The models fit the observations of C$_2$H$_2$ between a range of
times and locations after 0.1 Myr and inside of 18~au. 
Formation at early times ($<$ 0.1 Myr) is by freezeout of gas-phase
hydrocarbon ions and neutrals (C$_2$H$_3^+$, C$_2$H$_5^+$,
C$_3$H$_7^+$, and C$_2$H$_2$).  Cosmic ray photon destruction of
C$_2$H$_6$ ice also plays a role. Hydrogenation of C$_2$H is not a
major formation route at these times, but does contribute at
$t$ $>$ 0.1 Myr.
Destruction of C$_2$H$_2$ is by cosmic ray photons forming C$_2$H.

The model predictions for C$_2$H$_6$ fit the comet observations either
at early times ($<$ 1 Myr) inside of 15-20~au, or after 1 Myr at $R$
$>$ 20 au.  The formation process depends on the time and location,
with freezeout of gas-phase ions (mainly C$_3$H$_7^+$) dominating at
smaller radii and 
sequential hydrogenation of C$_2$H at larger radii.

\subsection{HCN}
HCN/H$_2$O is fairly constant over the disk with time and
radius.  Its initial abundance is 3.2\%, higher than observed in
comets (0.07 -- 0.5\%) and the comet data is only fit in regions where
desorption is efficient (i.e. near this molecule's snowline, around 6-7~au).  At larger
radii it is formed by freezeout of HCN or HCNH$^+$, or by reaction of CN and H on
the grains.  It is destroyed by cosmic ray photons, re-forming CN.

\subsection{OCS}

The molecular cloud mixing ratio of OCS is very low (OCS/H$_2$O =
0.02\%) and this species
mainly forms in the disk from the reaction of CO with sulphur atoms
(Equation~\ref{eq:ocs}). Destruction is by cosmic ray photons forming
CS or CO.  The disk OCS abundance
matches the comets either around its snowline, or  outside of 20~au
and after 10 Myr. 

\subsection{\label{sec:nh3}NH$_3$}
Ammonia ice forms efficiently in the molecular cloud from
hydrogenation of nitrogen atoms in the ice. Its initial abundance
is 4.7\%.  The highest value of NH$_3$/H$_2$O observed in a comet is  3.6\% in C/2012
S1, an Oort cloud comet.  For the other comets the ratio is less
than half of this.  Hence, some loss of NH$_3$ is required to match
the observations.  This is achieved in two regions on either
side of the CO snowline, at times later than 1 Myr, when NH$_3$ is destroyed by
cosmic ray photons forming NH$_2$ and NH.  NH$_2$ is likely to
react with H reforming NH$_3$, but NH does not do this.  Outside of the
CO snowline, NH reacts with CO to form HNCO
(Equation~\ref{eq:hnco}), whereas at smaller radii NH can
desorb. This results in a loss of NH$_3$.

%%%%%%%%%%%%%%%%%%%%%%%%%%%%%%%%%%%%%%%%%%%%%%%%%%%%%%%%%%%%%%%%%%%%%%
\section{\label{sec:lowcr}Low Cosmic Ray Model Results}
Cosmic rays are important drivers of the chemistry in the fiducial
model.  For
the interstellar medium an ionization rate of 1.3 $\times$ 10$^{-17}$
s$^{-1}$ is generally assumed, although
higher rates have been inferred from observations of H$_3^+$ in some
diffuse regions \citep{mccall03,indriolo07}.  For disks around young stars,
\cite{cleeves13} suggest that the ionization rate may be lower than
interstellar, with a value of 0.23 -- 1.4 $\times$ 10$^{-18}$
s$^{-1}$.   \cite{seifert21} show that the picture may be more
complicated, finding a very low value of $\zeta_{CR}$ $<$ 10$^{-20}$
s$^{-1}$ in the inner 100 au of IM Lup, and 10$^{-17}$ s$^{-1}$
outside of this. 

\begin{figure}

\includegraphics[trim=275 20 20 20, clip, width=1.1\linewidth]{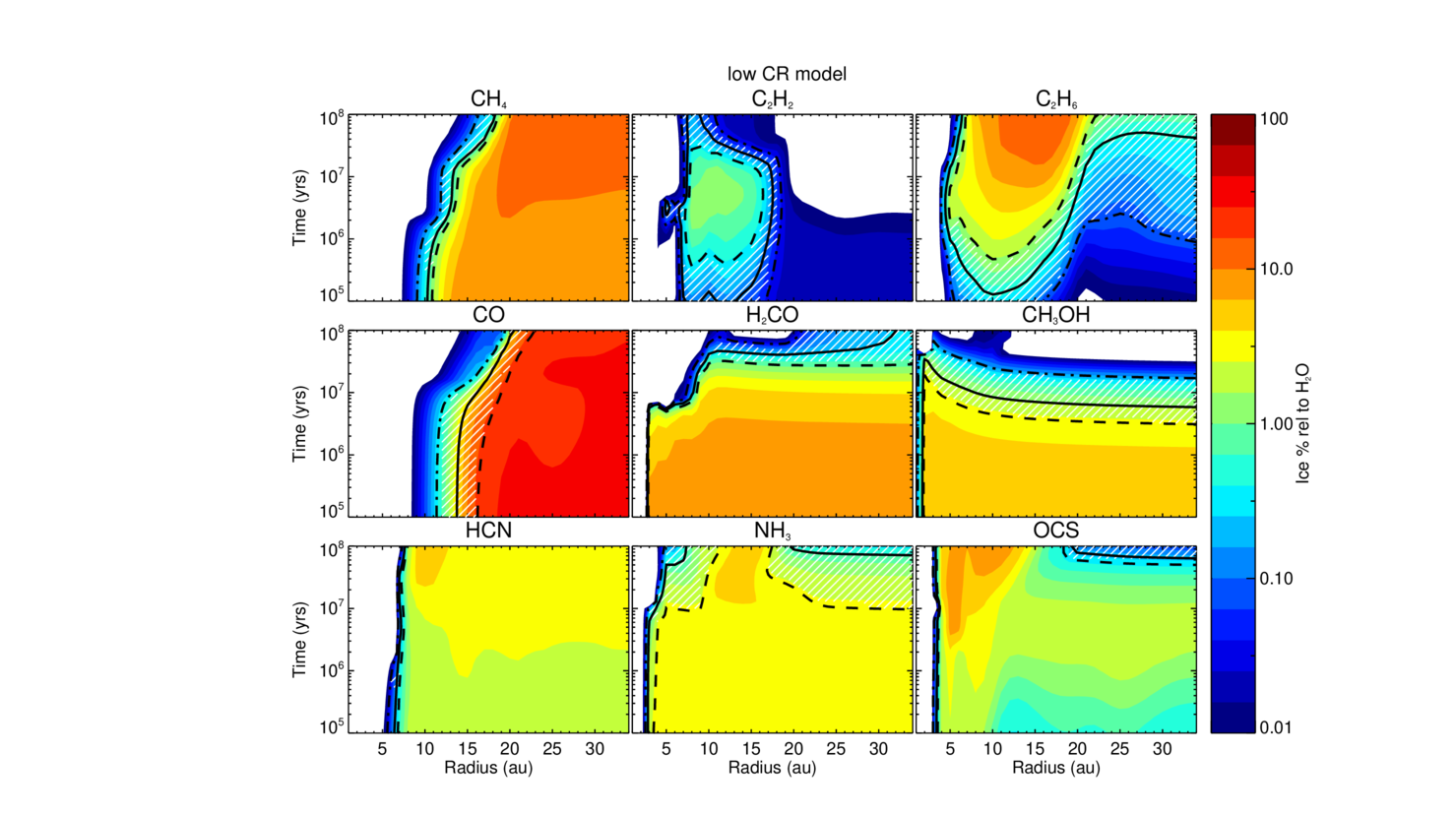}
\caption{\label{fig:ratio_lowCR}Calculated abundances (filled
  colored contours) from the  model with low cosmic ray flux compared to
  the observed abundance ranges in all comets in Appendix 2.} 
\end{figure}

To examine what effect a lower cosmic ray ionization rate
might have on the midplane ice abundances we ran a second model with
$\zeta_{CR}$ = 1.3 $\times$ 10$^{-18}$ s$^{-1}$, a factor
of 10 lower than the flux in our fiducial model (Figure~\ref{fig:ratio_lowCR}).  Longer timescales are required to
match the observed range for those species whose chemistry is
driven by cosmic ray
processing of the ices.  For example, cosmic rays destroy the CO at 5
-- 10~Myr, rather than 0.5 -- 1~Myr as in the fiducial model. 
The model H$_2$CO now matches
the observations only at $t$ $>$ 20 -- 100 Myr and CH$_3$OH between
5 and 50 Myr.  Unlike the fiducial model a low cosmic ray ionization
rate cannot fit the entire range of the NH$_3$
observations at $R$ $>$ 20~au, and only the lowest part of the
observed OCS range can be accounted for.
However, this 
model is still able to match the observed range of mixing ratios
for the other molecules, albeit it at later times than in the
fiducial model.

%%%%%%%%%%%%%%%%%%%%%%%%%%%%%%%%%%%%%%%%%%%%%%%%%%%%%%%%%%%%%%%%%%%%%%
\section{\label{sec:reset}`reset' Model Results}

 Following other studies of protosolar nebula composition we also consider a model where any molecular cloud composition is wiped out by the formation of the disk, and the disk chemistry starts with atomic abundances.  This could occur during the infall process, e.g. \cite{visser09}, or by energetic events in the early solar system.  There is evidence for some degree of `reset' in the inner solar system from studies of chondrules and CAIs, e.g. \cite{triniquier09}, but whether this would extend out to the comet formation region is uncertain.

\begin{figure}

\includegraphics[trim = 275 20 20 20, clip, width=1.2\linewidth]{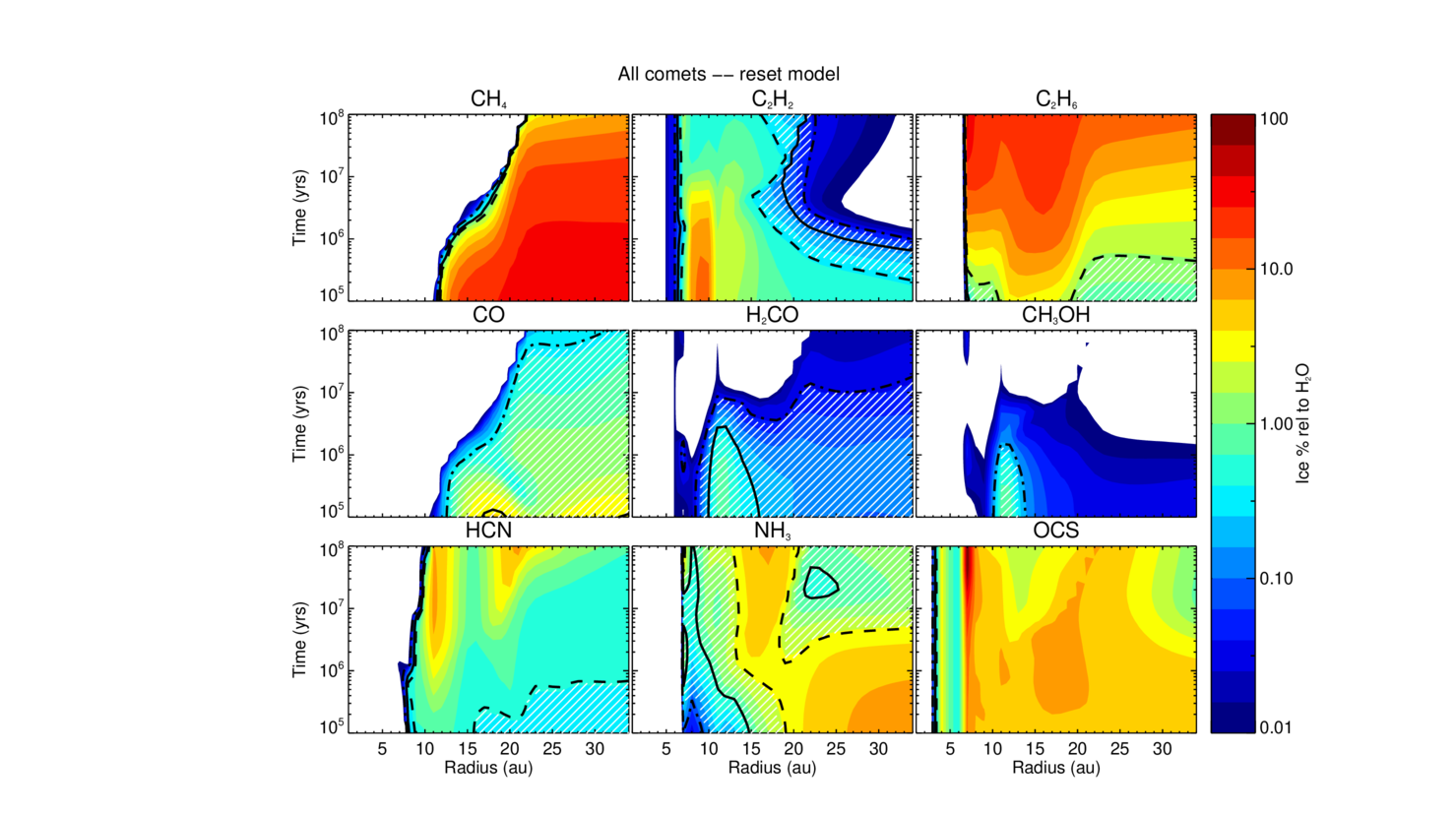}
\caption{\label{fig:reset}Predicted molecular distributions for the
  `reset' abundance inputs (colored contours).  Markings are the same as
  in Figure~\ref{fig:ratio_all}. }
\end{figure}

The abundances for the `reset' model are are given by 
Table~\ref{tab:init}.  The resulting disk composition is shown Figure~\ref{fig:reset}.   The magnitude and distribution of the mixing ratios are quite different to the fiducial model

The CO distribution is similar to the fiducial model but much lower. For most of the disk the ratio of CO/H$_2$O $\sim$ 1\%.  It does reach $\sim$ 5\% at time earlier than 0.1 Myr but this is much lower than the ratio seen in many comets. CO ice is converted into CO$_2$, HNCO and OCS. H$_2$CO and CH$_3$OH, molecules that derive from the hydrogenation of CO, are both under abundant compared to the fiducial model.  While H atoms are abundant at the start of the disk model, there is an activation barrier to hydrogenation of CO.  The warmer temperatures in the disk compared to the molecular cloud model help to overcome this. They also mean that the residence time of H atoms on the grains is reduced and that other reactions of CO (with NH and S) are also faster than in the molecular cloud and hence the formation of H$_2$CO and CH$_3$OH is less efficient. The `reset' model can account for the lowest mixing ratios of these molecules in comets, but it cannot account for the full range of observations. 

C$_2$H$_6$ matches the observations on either side of the CO snowline.  This distribution resembles that seen in Figure~\ref{fig:ratio_all} but the match to observations occurs at earlier times.  C$_2$H$_2$ on the other hand is much more abundant than in the fiducial model with ratios of $>$ 10\% at 8-10 au.  The fiducial model matches the observations of C$_2$H$_2$ around the CO snowline, but in the `reset' model the match is either at $R$ $>$ 15 au, or at 6-10 au, depending on the time. 

The `reset' model can at least partially match the HCN observations at larger radii ($R$ $>$ 15 au) than was possible in the fiducial model.  The NH$_3$ observations are matched either at late times ($t$ $>$ 1Myr) and outside of 20 au, or at smaller radii at all times.  The model OCS only agrees with the observed range around its snowline.

Overall, the `reset' model is unable to account for the complete range of observed abundance ratios.

\section{\label{sec:families}Comparison of observations in JFCs and OCCs}

The results above show that the fiducial model can account for the complete
range of abundances in Table~\ref{tab:av},
as reported in DR16 and \cite{saki20}.  We now consider whether
there are any systematic differences between the Jupiter-family and
Oort cloud comets.  One caveat to this discussion is that the observed
abundance ranges we use are based on a relatively small number of
measurements and so are unlikely to represent the complete range of
possible abundance ratios in
comets.  
In particular, the number of measurements in JFCs is quite
small. Therefore any conclusions regarding the differences between the families are tentative
because they could be a
result of the small  sample size.
\begin{figure}
  \centering
  \includegraphics[width=0.44\linewidth]{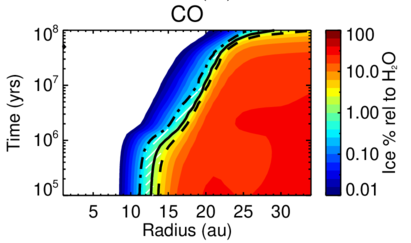}
  \includegraphics[width=0.44\linewidth]{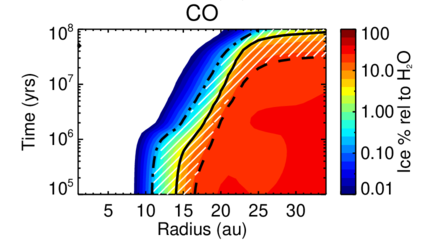}\\
  \includegraphics[width=0.44\linewidth]{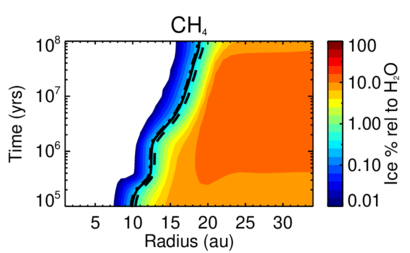}
  \includegraphics[width=0.44\linewidth]{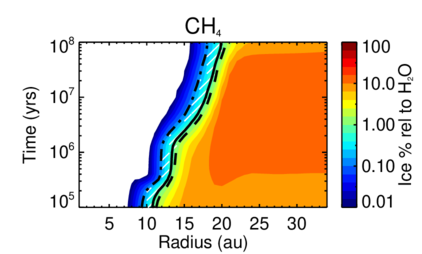}\\
  \includegraphics[width=0.44\linewidth]{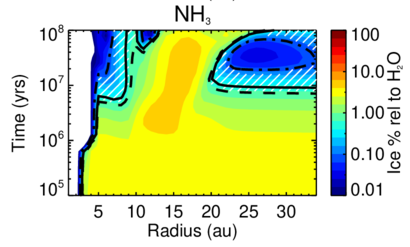}
  \includegraphics[width=0.44\linewidth]{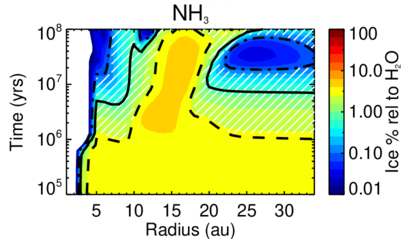}\\
  \includegraphics[width=0.44\linewidth]{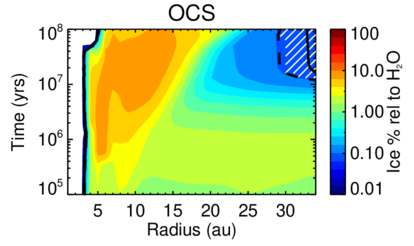}
  \includegraphics[width=0.44\linewidth]{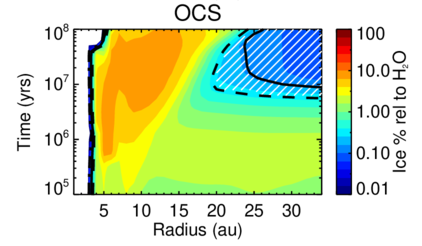}\\
  \includegraphics[width=0.44\linewidth]{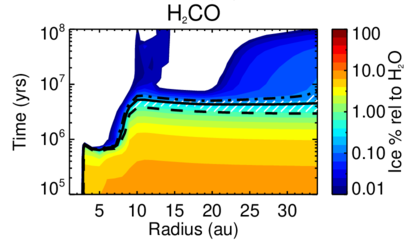}
  \includegraphics[width=0.44\linewidth]{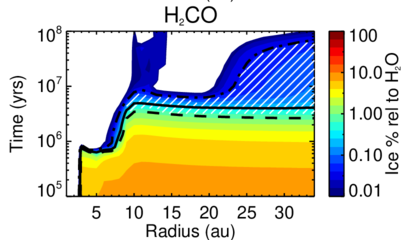}\\

  \caption{\label{fig:ratio_jfc}Average and range of the JFC 
    ({\it left}) and OCC ({\it right}) family observations, showing
    the range of disk times and locations which fit the
   observed mixing ratios relative to H$_2$O from Table~\ref{tab:av}. Lines and symbols are as in
    Figure~\ref{fig:ratio_all} for each comet family.}
  \end{figure}
\addtocounter{figure}{-1}
\begin{figure}
  \centering
  \includegraphics[width=0.44\linewidth]{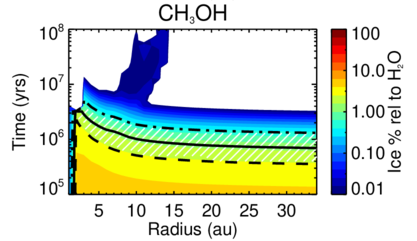}
  \includegraphics[width=0.44\linewidth]{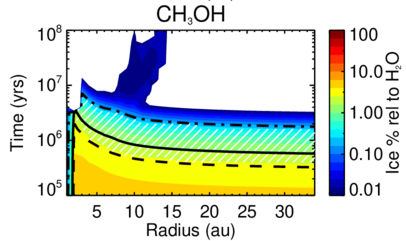}  \\
  \includegraphics[width=0.44\linewidth]{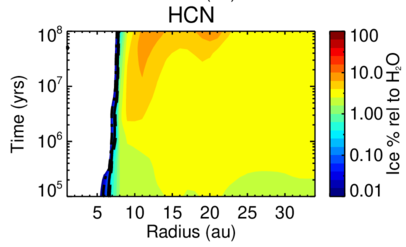}
  \includegraphics[width=0.44\linewidth]{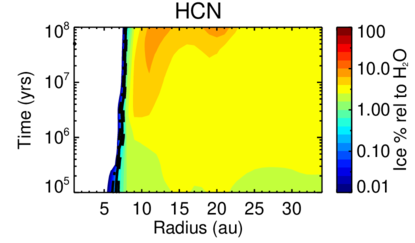}  \\
  \includegraphics[width=0.44\linewidth]{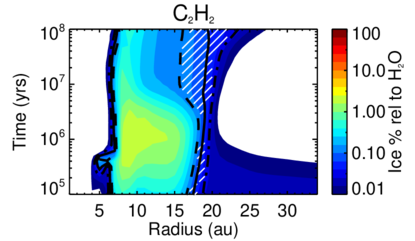}
  \includegraphics[width=0.44\linewidth]{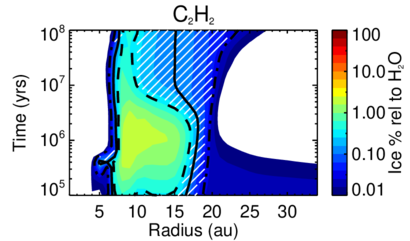}  \\
  \includegraphics[width=0.44\linewidth]{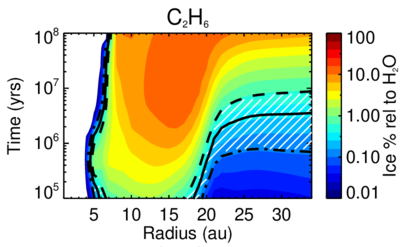}
  \includegraphics[width=0.44\linewidth]{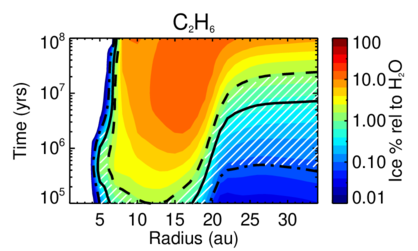}  \\
\caption{cont}
\end{figure}
Comparing the models to the observed ranges suggests the differences
between the JFCs and OCCs are small (Figure~\ref{fig:ratio_jfc}).  The
molecule with the 
biggest difference  is CO whose abundances
span a greater range in OCCs than JFCs, corresponding to a
larger range of radii in the model protosolar nebula. This difference is
enhanced if the observed upper limit is taken to be the largest
individual measurement in Hale-Bopp (41\%) rather than the weighted
average of 26\% in this comet.  With this higher value the outer radius of models
that fit the OCC range is now 35~au rather than 18~au at
0.1 Myr.  NH$_3$ also shows
some differences between the two families, with a larger range of
model times required to fit the OCC observed range.  A similar effect is seen
for H$_2$CO and C$_2$H$_6$. For other molecules
though, the combinations of location and time that fit the observed
range show little difference between the two families.

\cite{ahearn12} used observations of CO, CO$_2$ and H$_2$O to
investigate where JFCs and OCCs formed.  They found that JFCs and OCCs
formed in overlapping regions, with JFCs on average slightly closer to
the Sun.  This is consistent with our results if the highest CO/H$_2$O
ratio observed in Hale-Bopp is taken into account.  In particular, the
narrower range of CO mixing ratios in JFCs is fit by the model disk's
ices only around the CO snowline, whereas for OCCs the larger range
of CO mixing ratios could allow some to form as far out as 35~au.  However,
this is the opposite of the classical picture where JFCs form beyond
Neptune and OCCs in the region of the giant planets before being
scattered to their current reservoirs \citep{rickman10}.

Alternatively, JFC abundances may reflect their time spent as Centaurs
orbiting among the outer planets.  Here, the outer kilometer or so of
the nucleus can be heated by sunlight to 80 or 90~K \citep{gl16},
sufficient for some of the more-volatile species to be partially lost
\citep{lauck15}.  Such losses may account for the JFCs' generally
lower CO abundances, if today's JFCs retain some of the
thermally-processed material.  However, it is unclear whether there is
a thermally processed layer, and if so how much of the present-day
outgassing comes from this material rather than the less-processed
interior of the nucleus.  Enough perihelion passes near the Earth's
orbit would remove the thermally-processed layer, revealing the bulk
nucleus.  This appears to be the case for 67P, where the presence of
N$_2$ suggests that any thermally processed material has already been
lost \citep{rubin15}.

\section{How did comets  obtain their compositions?}

\subsection{\label{sec:1rad}Assembly from material at a single time and location}

While the previous section has shown that the fiducial and low cosmic ray models can match the
observed range of each molecule {\it individually} the question now is
whether they can provide an explanation of the complete composition of
each comet as a whole.  
To determine how well 
our models account for the {\it overall} comet compositions we compare them to a 
subset of individual comets from our sample, consisting of 
the JFCs 103P/Hartley 2 
and 9P/Tempel 1, and the OCCs C/1999 H1 (Lee),
C/2009 P1 (Garradd), C/2013 R1 (Lovejoy), C/2007
N1 (Lulin), C/2012 S1 (ISON) and C/2004 Q2 (Machholz).  Each of these comets has observations for all eight molecules
discussed in DR16, although only upper limits are available for
 CH$_4$ in Hartley 2
and for H$_2$CO and C$_2$H$_2$ in Lovejoy.  For ISON
we use the data listed in Table 3 of DR16 as coming from $\geq$ 0.83
au since this has the most complete set of molecules.  Where
upper limits only are available we assume the mixing ratio to be half
of the upper limit.  In addition, OCS observations are available
for comets Lovejoy, Garradd, ISON, and Lee \citep{saki20}.   For this analysis we use the fiducial model, since the `reset' model cannot match the complete range of observed mixing ratios.

In addition to the individual comets we also analyze the average
composition for JFCs and OCCs from Table~\ref{tab:av}.  Although no comet has this
exact composition, the averages do serve as a means of determining if
there are any systematic differences in the times and locations at
which the different families might have formed. 

To assess whether the models provide a good fit to the data we use the
$\chi^2$ test.
For each combination of location and time for our fiducial model
we calculate the value of $\chi^2$ from
\begin{equation}
\label{eq:chi}
\chi^2 = \sum\limits_{i=1}^n \frac{(O_i - M_i(r,t))^2}{M_i(r,t)}
\end{equation}
where $O_i$ is the observed mixing ratio relative to water ice of molecule $i$ and $M_i(r,t)$ is the mixing ratio $i$ relative to
water ice for a model at radius, $r$ and time, $t$.  A good fit of the
models to the observations at a 95\% confidence level requires
$\chi^2$ $<$ 14.07 for comets without OCS observations and $\chi^2$
$<$ 15.5 for those with measured OCS.  Note that \cite{eistrup19} (hereafter E19) carried out a similar analysis but used
a slightly different version of $\chi^2$ since they were comparing the
logarithm of the observations and models.  Use of their formulation
results in some differences in the calculated $\chi^2$ values but does not
change our overall conclusions.

\begin{figure}
   \includegraphics[width=0.35\linewidth,trim=125 0 270 0,clip]{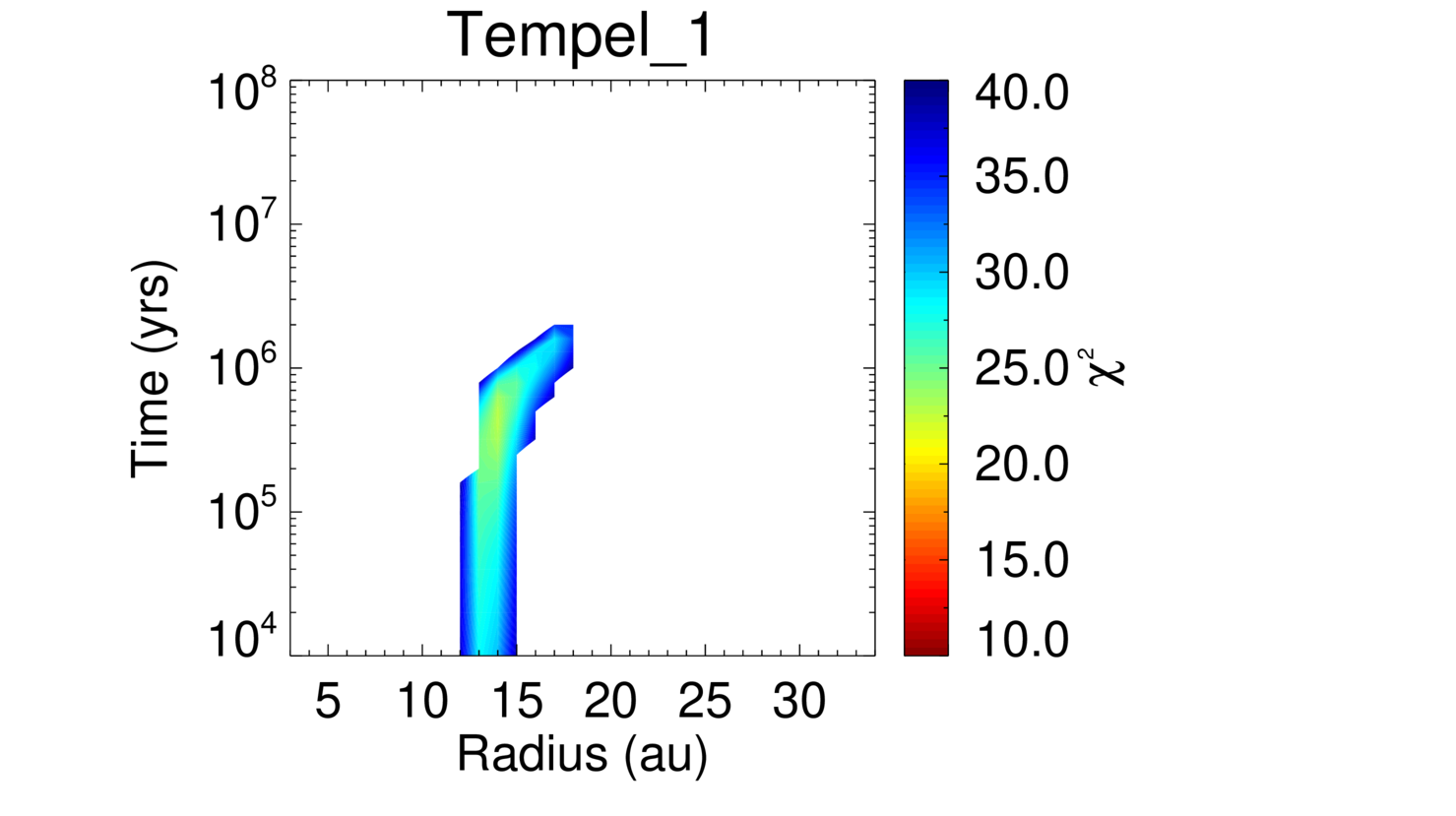}
   \includegraphics[width=0.35\linewidth,trim=125 0 270 0,clip]{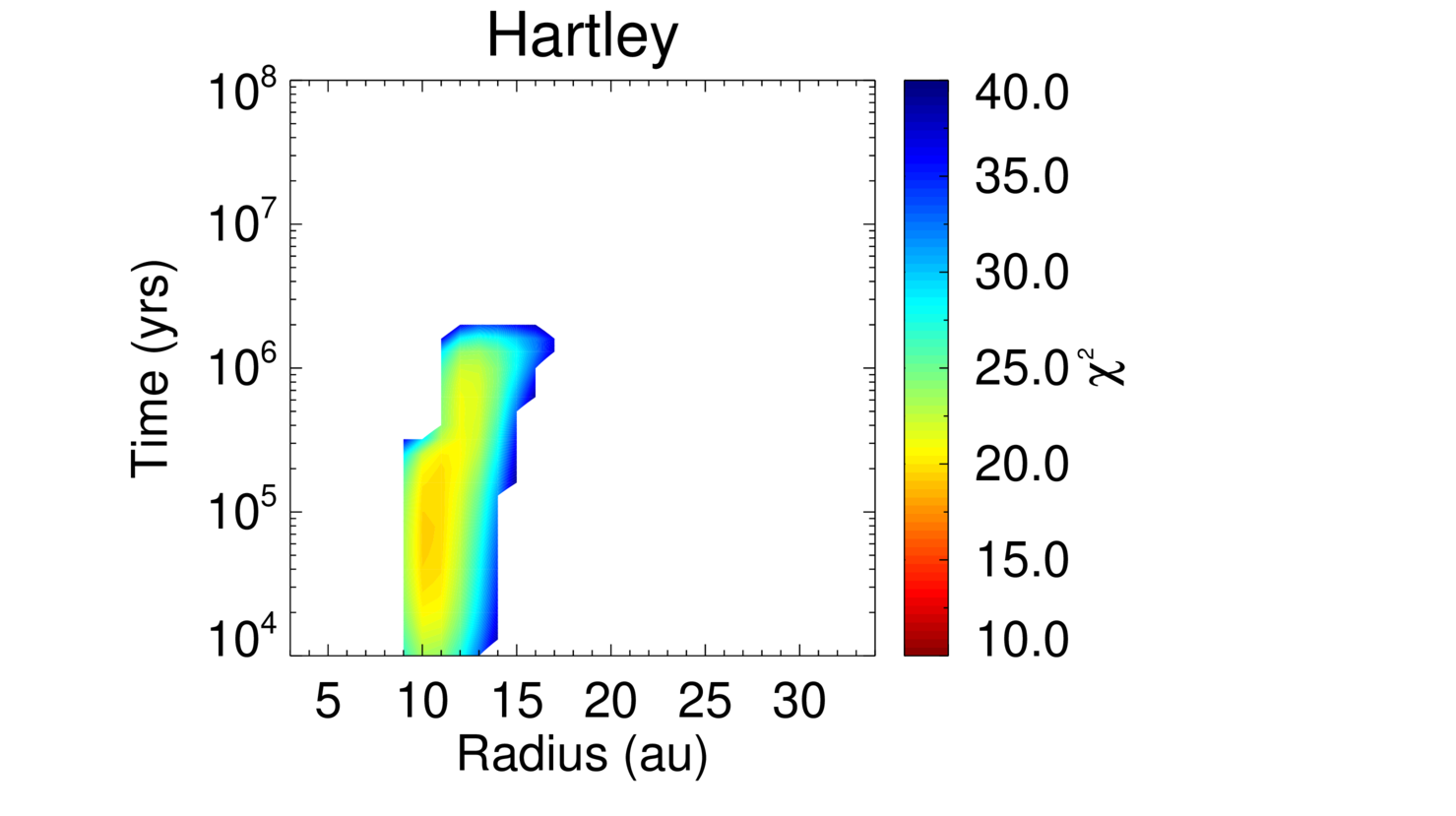}
   \includegraphics[width=0.35\linewidth,trim=125 0 270 0,clip]{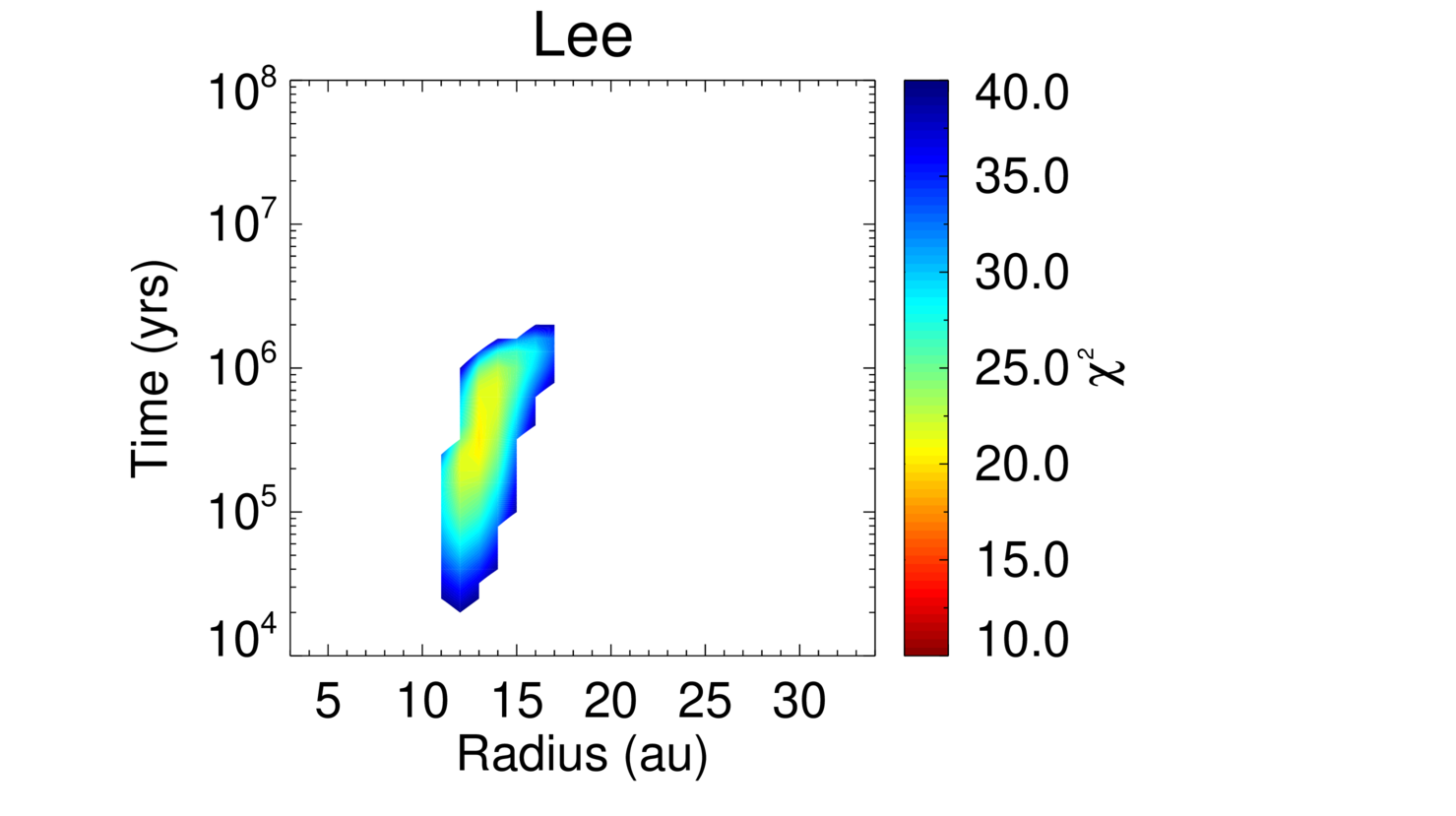}
   \includegraphics[width=0.35\linewidth,trim=125 0 270 0,clip]{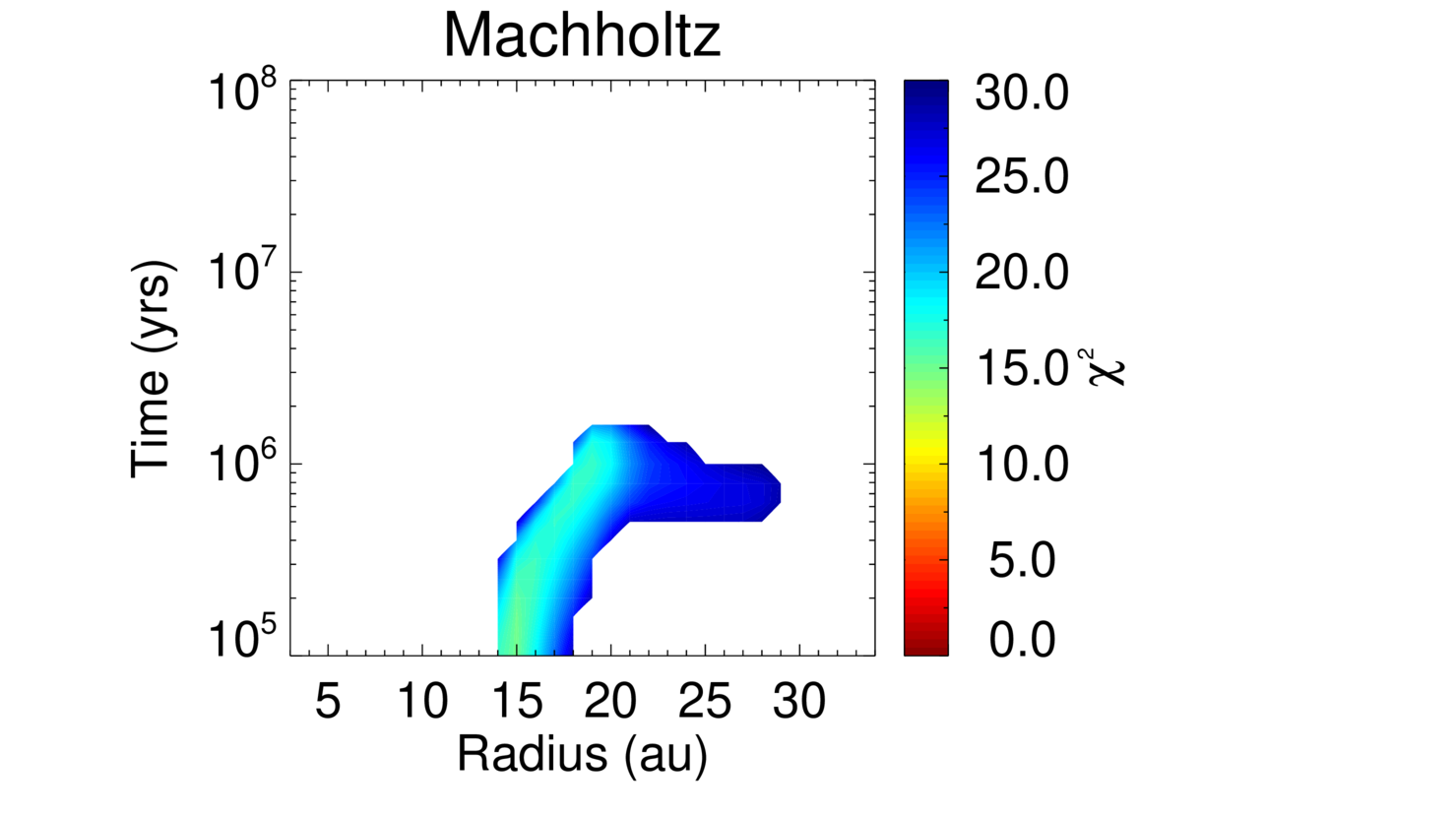}
   \includegraphics[width=0.35\linewidth,trim=125 0 270 0,clip]{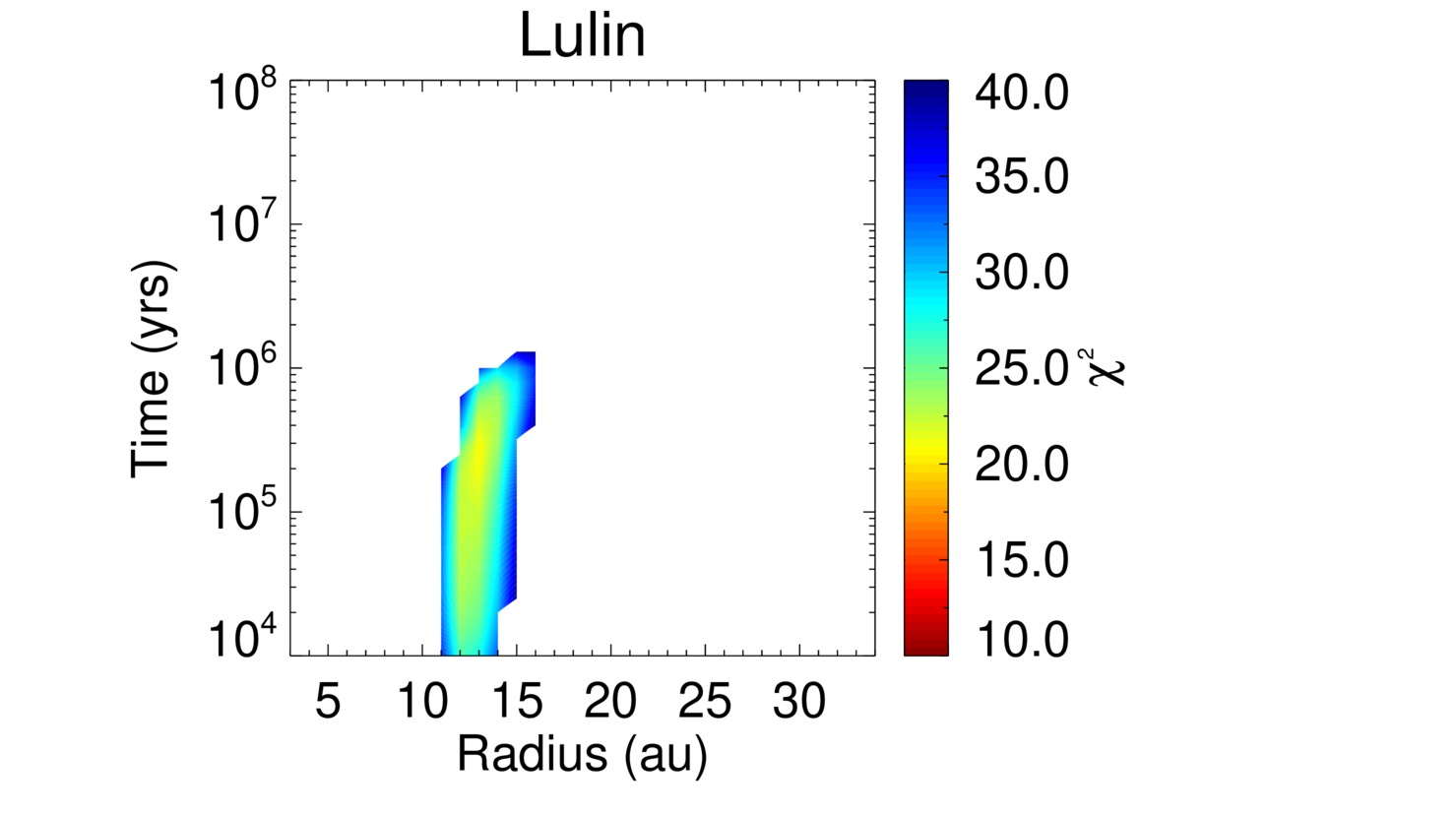}
   \includegraphics[width=0.35\linewidth,trim=125 0 270 0,clip]{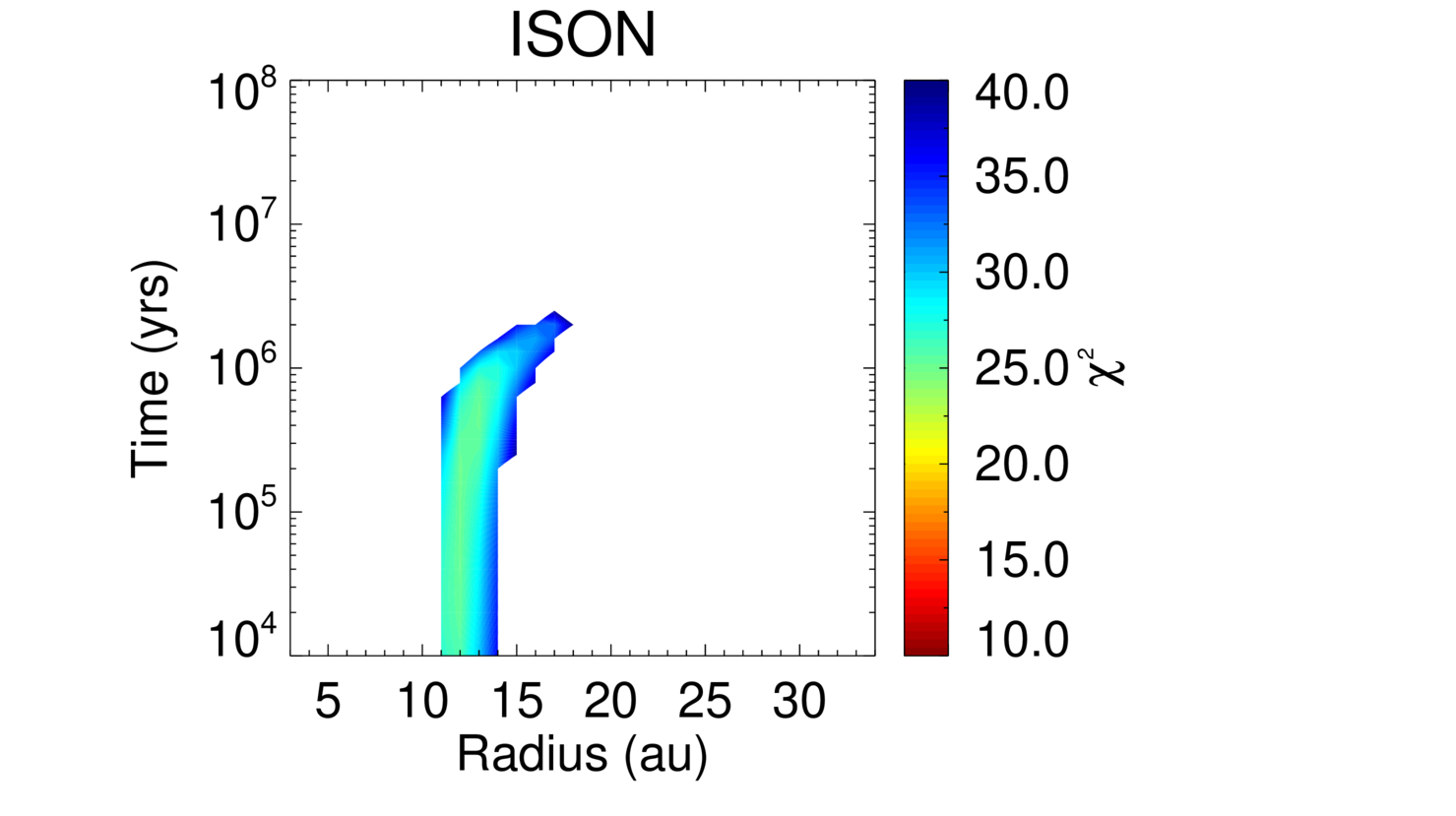}
   \includegraphics[width=0.35\linewidth,trim=125 0 270 0,clip]{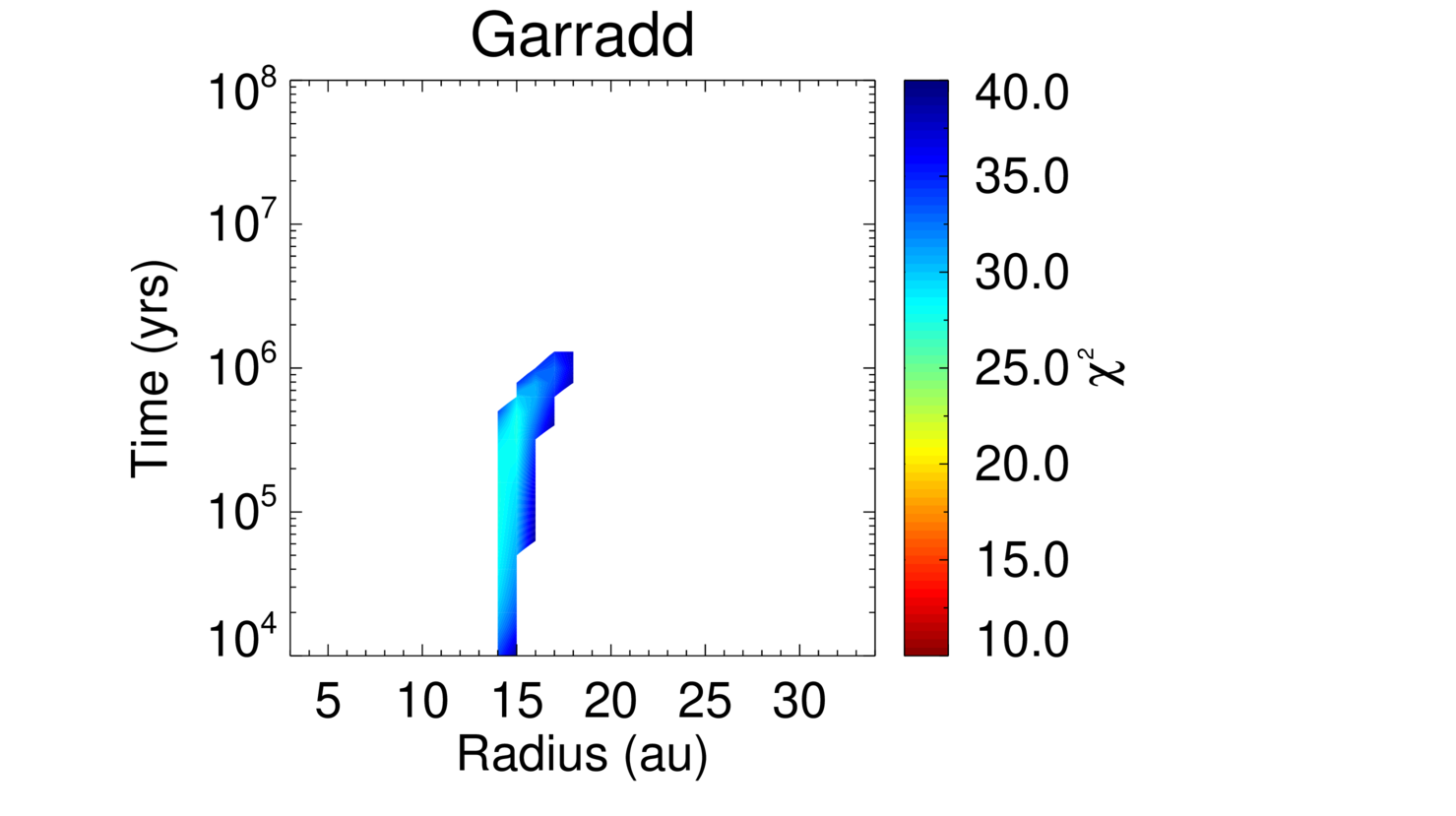}
   \includegraphics[width=0.35\linewidth,trim=125 0 270 0,clip]{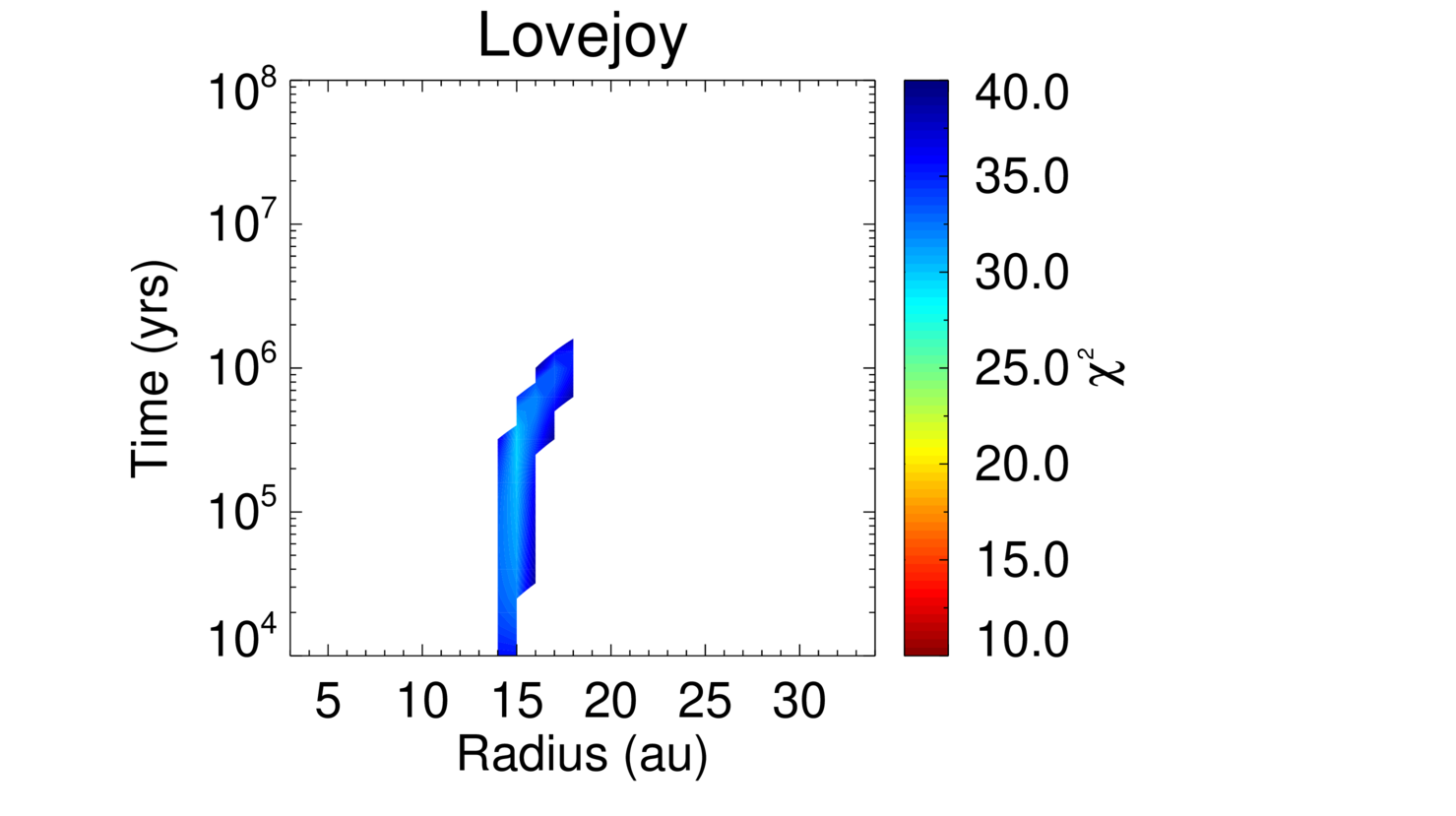}
   \includegraphics[width=0.35\linewidth,trim=125 0 270 0,clip]{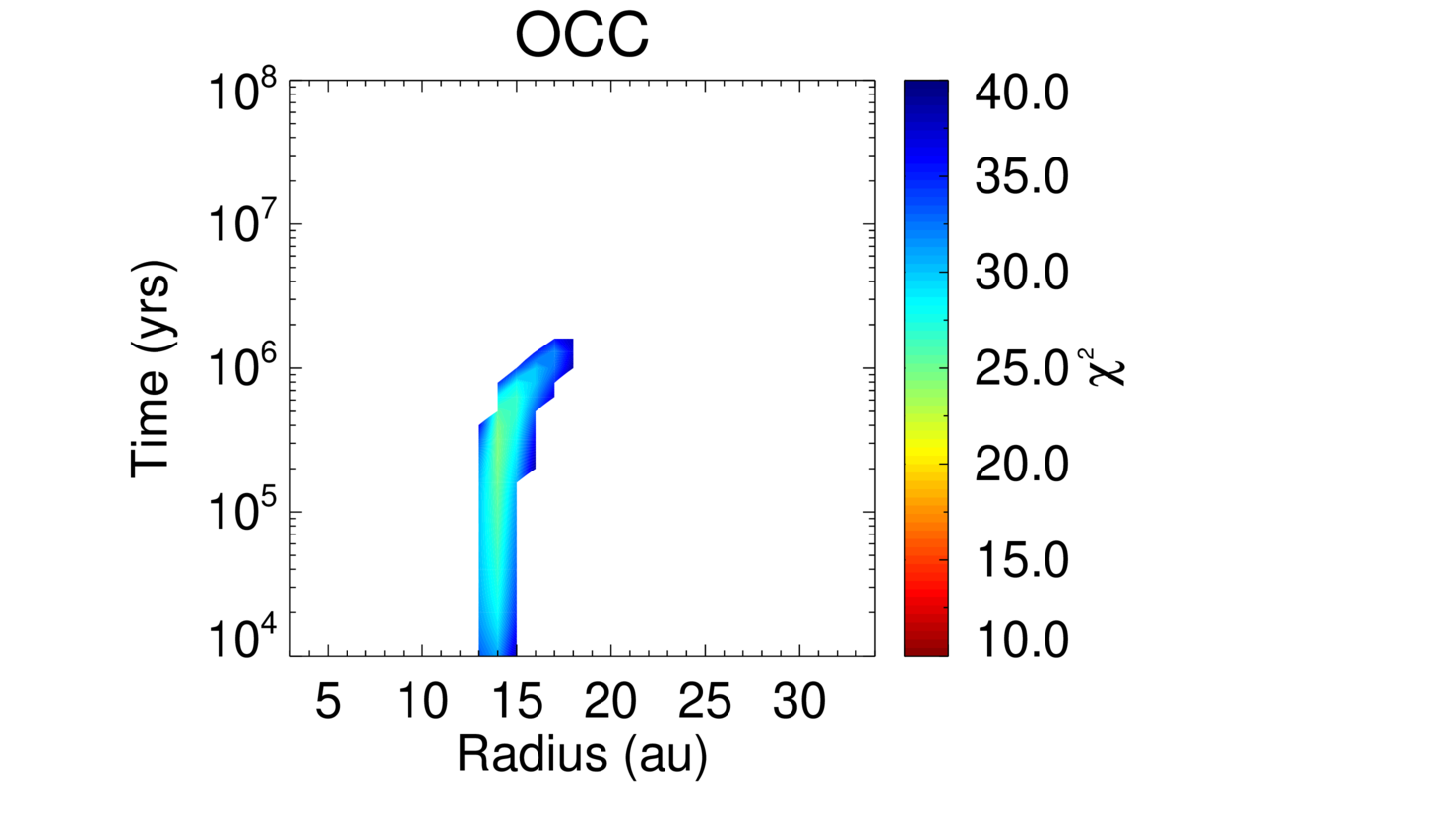}
   \includegraphics[width=0.35\linewidth,trim=125 0 270 0,clip]{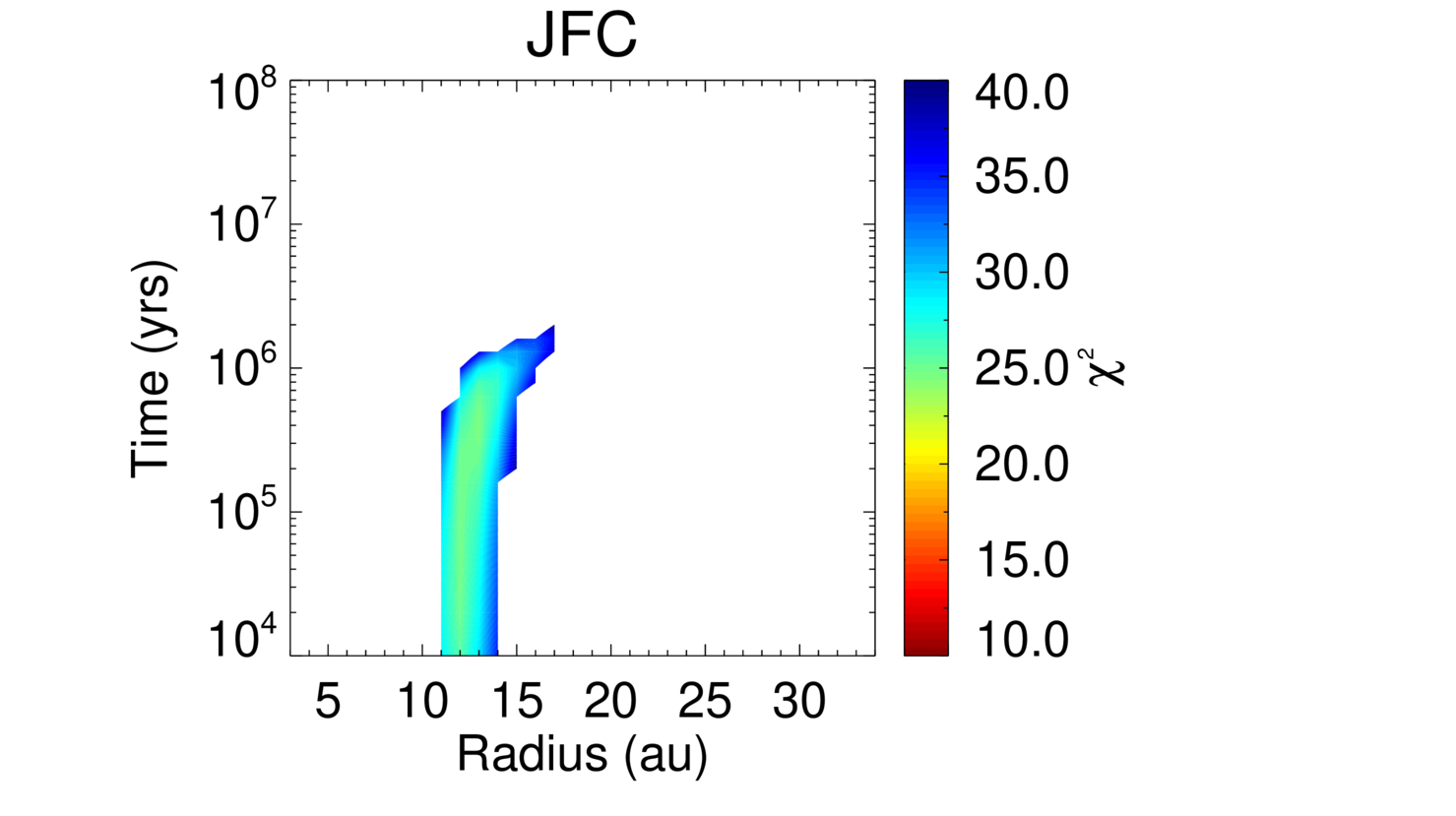}
\caption{\label{fig:chi2}Plots of $\chi^2$ calculated by
  Equation~\ref{eq:chi} comparing observations of individual comets
  with models. Also shown are $\chi^2$ calculated for the average
  compositions of the JFC and OCC families. }
\end{figure}

The calculated $\chi^2$ obtained by comparing the observations with the fiducial model are  shown in Figure~\ref{fig:chi2}. The
lowest $\chi^2$ are found around the CO snowline in agreement with the
results of E19.  The location and time for the best fit for each comet are listed in
Table~\ref{tab:1fit}. However,  in
each case the lowest calculated $\chi^2$ is above the limit required
for a good fit at the  95\% confidence level. 

 The lowest $\chi^2$ found for the average JFC and OCC compositions are
in similar locations, with  a radius of 13 au for JFCs and 14 au for
OCCs.  
For the individual comets the locations range from 10 au (for Comet Hartley 2)
to 15 au for Lovejoy and Garradd, and the times from 0.08 Myr (Comet Hartley 2)
 to 0.5 Myr (for comets Tempel 1 and ISON).  It is perhaps not
 surprising that the $\chi^2$ suggests that the models at a single location
 and time are not a good fit to the overall composition of the comets, as Figure~\ref{fig:ratio_all} shows that the molecules come from
 different (non-overlapping) regions in the disk.  For example, the predicted HCN mixing
 ratios only cover the observational range close to its iceline around 6 au,
 much closer to the star than the location of the lowest $\chi^2$ models. 

In the literature, good agreement between individual observations and
 astrochemical models is often taken to be a factor of 10.  This allows for errors in
the observational data, as well as in the models themselves. The final column in
 Table~\ref{tab:1fit} lists the molecules for which the predicted
 abundances of the best fit model deviate from the observations by
 more than this.
None of the models can fit HCN, and in several
comets
 H$_2$CO, C$_2$H$_2$ and C$_2$H$_6$ are not matched either.  For the
 average compositions  the models cannot
fit HCN and H$_2$CO, with C$_2$H$_2$ and C$_2$H$_6$ also in poor agreement with the average JFC values.  

Using Hartley 2 as  an example, we look at the size of the discrepancy
between observations and models at the location of
the lowest $\chi^2$.  The observed mixing ratio for HCN is 0.24\%,
but at $R$ = 10 au and $t$ = 0.08 Myr the model predicts HCN/H$_2$O =
3.2\%, a difference of a factor of 13.  The difference in the observed
and predicted mixing ratios of H$_2$CO at this location is even
higher -- a factor of 64.  
The predicted abundances from a single location and time in the
protosolar nebula therefore do not provide a good match to the overall
composition of our comets.  We now investigate a mechanism to improve the
agreement.

\begin{deluxetable}{lllll}
  \tablecaption{Best fit models from a single radius and time to the
    observational data.  Molecules listed are those which cannot be
    fit by these models to within a factor of 10 of the observed
    value.Good fits are provided by models with $\chi^2$ $<$ 14.07 for comets with eight observations (8P/Temple 1, 103P/Hartley 2, C/1999 H1 (Lee), C/2004 Q2 (Machholz) and C/2007 N1 (Lulin)), and $<$ 15.5 for those with nine observations (C/2009 P1 (Garradd), C/2012 S1 (ISON) and C/2013 R1 (Lovejoy)). \label{tab:1fit}}
  \tablewidth{0pt}
  \tablehead{
    \colhead{Comet} & \colhead{$\chi^2$} & \colhead{R (au)} &
    \colhead{time (Myr)} & 
\colhead{Molecules not fit by lowest $\chi^2$ model} } 
\startdata
  9P/Tempel 1 & 23.2  & 14 &  0.50   & HCN, NH$_3$, H$_2$CO, C$_2$H$_6$ \\
  103P/Hartley 2 & 19.9 & 10 &  0.08   & HCN, H$_2$CO \\
  C/1999 H1 Lee  & 20.3 & 13 & 0.32     & HCN, H$_2$CO\\
  C/2004 Q2 Machholz  & 24.3 & 14 & 0.04 & HCN, NH$_3$, H$_2$CO, CH$_4$,
  C$_2$H$_2$, C$_2$H$_6$\\
  C/2007 N1 Lulin & 21.1 & 13 & 0.25       & HCN, NH$_3$, H$_2$CO, C$_2$H$_2$\\
  C/2012 S1 ISON & 20.1 & 13 & 0.50  & HCN, C$_2$H$_6$ \\
  C/2009 P1 Garradd & 28.1 & 15 & 0.40   & HCN, H$_2$CO, C$_2$H$_2$\\
  C/2013 R1 Lovejoy & 30.3 & 15 & 0.25   & HCN, NH$_3$, H$_2$CO, C$_2$H$_6$\\
 Average OCC & 24.9 & 14 & 0.32 & HCN, H$_2$CO\\
 Average JFC & 25.1 & 13 & 0.40 & HCN, H$_s$CO, C$_2$H$_2$, C$_2$H$_6$\\
  \enddata
\end{deluxetable}

\subsection{\label{sec:2rad}Disk mixing -- combining material from two locations}

The distribution of molecules seen in Figure~\ref{fig:ratio_all} suggests
that combining material from different radii might result in a closer
match to the observed comet compositions.  Disks experience widespread turbulence, likely resulting in radial and vertical mixing throughout.  Both mechanisms
would bring material processed in warmer and/or higher UV regions,
where more volatile species such as CO have been lost, into
regions with CO-rich ices where they could combine to form protocomets.
We explore this possibility with a toy model that combines material
from inside and outside the CO iceline.  Although not a physically complete solution, this model explores a potential means of explaining the range of thermal histories required to explain material incorporated into cometary nuclei. We note that this is not
`the solution' and a more complex exploration of parameters, such as 2-D (vertical and radial) mixing is likely needed and will be the focus of future work.  

There is observational evidence for the idea that inner disk material was incorporated into comets.  The presence of crystalline silicates in some comets \citep{bregman87,crovisier96,zolensky06} indicates that they include material processed at high temperatures from close to the star into their nuclei. Note that temperatures high enough to form crystalline silicates are not required for the scenario we are exploring here.

We consider a scenario where grains from two radii, $R_1$ and $R_2$
are combined. Since comets contain CO and N$_2$ we assume that they
must form outside of the CO snowline, and that grains from inside the
snowline are transported outwards to this location.  To achieve the
observed composition we therefore assume that material is transported
from the inner disk (where the models can fit HCN) to the outer
disk without any changes to its composition. We contend that this is a
reasonable assumption since (a) mixing from warm to cold regions will
not induce any desorption, and (b) condensation of new material is
unlikely since any species that can be condensed at $R_2$ will already
have done so on grains that are already present.  The inner disk ice
composition is therefore likely to be preserved as it travels
outwards to the comet formation zone. The travel time from
$R_1$ to $R_2$ may allow for chemistry to alter the composition, or
for gas traveling with the grains to  condense, but the complex modeling
to simulate these effects is beyond the scope of this manuscript.  
More sophisticated models incorporating coupled transport and chemistry 
are required to clarify  whether these mechanisms play a role in determining ice
compositions and will be addressed in future works.

The fiducial models discussed in Section~\ref{sec:fiducial} provide 
molecular abundance relative to H$_2$O as a function of time and 
radius in the disk. We use these results 
to determine whether material from two different locations can be combined 
to provide better agreement with the observations than those from a single time
and radius. For each model timestep between 10$^4$ and 10$^8$ yrs
we combine the abundance predictions from two radii, $R_1$ and $R_2$, in 
different proportions and determine whether the models can fit the data. 
We assume that the two radii come from inside ($R_1$) and outside ($R_2$)
the CO snowline.  $R_1$ is chosen from 3 -- 10 au, and $R_2$ from 10 -- 25 au, in increments of 1 au. 
We take each value of $R_1$ in turn and combine it with each possible value 
of $R_2$, with proportions of $R_1$ from 0.05-0-0.95 (in steps of 0.05) of 
the total comet composition.  
 The combined ice composition is then 
given by
\begin{equation}
  \hbox{combined composition} = f_{R_1}
  \frac{n_{R1}(X)}{n_{R1}(\hbox{H$_2$O})}+ (1-f_{R_1})
  \frac{n_{R2}(X)}{n_{R2}(\hbox{H$_2$O})}
  \label{eq:frac}
\end{equation}
where f$_{R_1}$ is the fraction of the total comet composition from
$R_1$, and n$_{R1}$(X) and n$_{R2}$(X) are the ice
abundances of X at $R_1$ and 
$R_2$, respectively.  In this way we can make a grid of compositions that cover all possible combinations of $R_1$ and $R_2$, resulting a total of 104386 combined compositions. We compare each with the comet
observations and
calculate the $\chi^2$.  To check that the models are able to produce
a reasonable agreement with each molecule's observed values, we also
determine whether all 8 (or 9 if OCS is observed) of the measured mixing ratios match the
model predictions to within a factor of 10. A good fit is 
therefore defined by (a) $\chi^2$ $<$ 14.07 (15.5) for 8 (9) observed
molecules (criteron C1) and (b) the models matching the observations
to within an order of magnitude (criteron C2).

\begin{figure}
\includegraphics[width=0.45\linewidth]{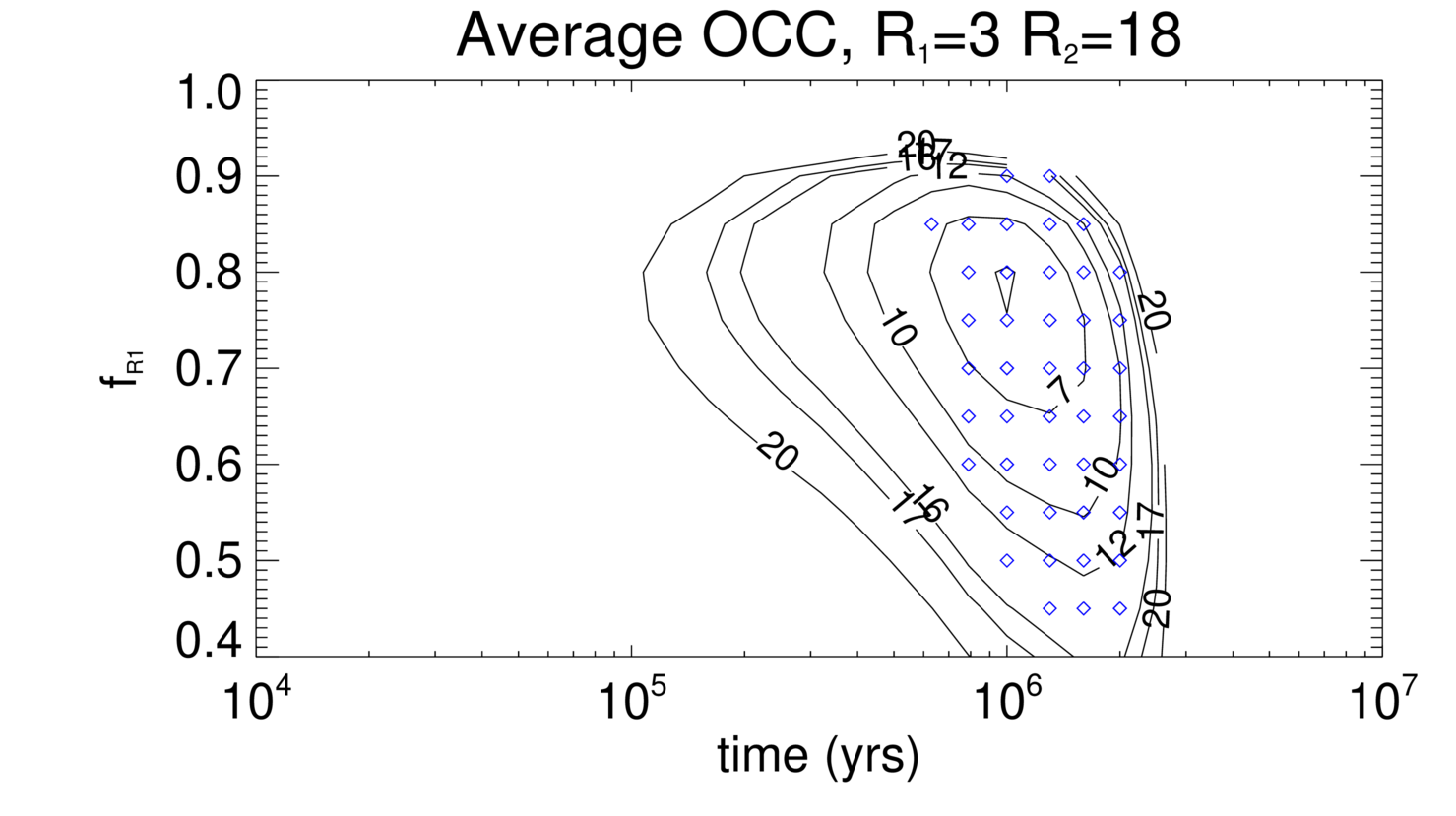}
\includegraphics[width=0.45\linewidth]{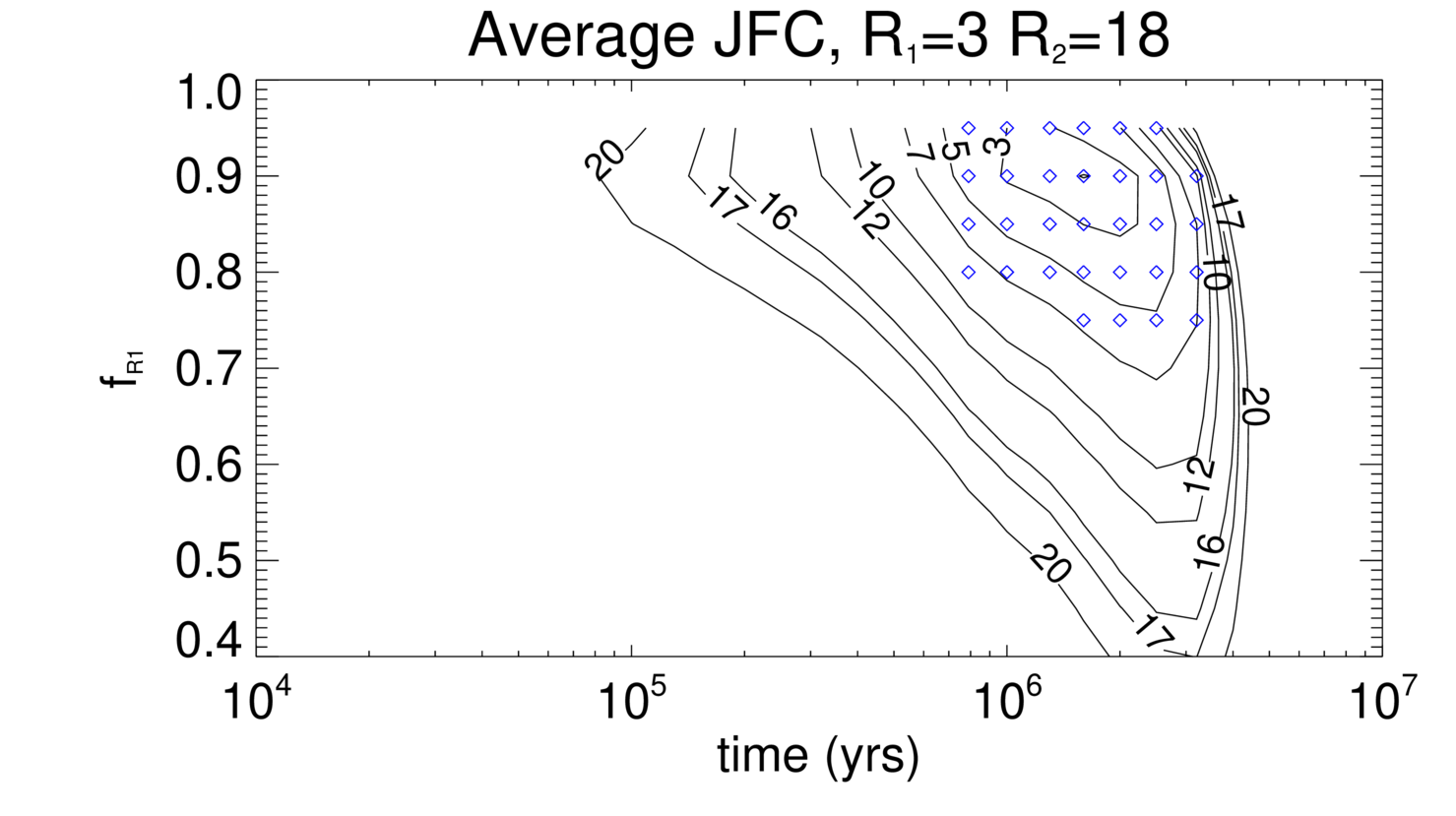}
\caption{\label{fig:2rad_av}The best fit models using combinations of two
  radii (one from inside the CO snowline (R$_1$), and one from outside it (R$_2$) to
  the average comet compositions. $\chi^2$ was calculated for all models for all
  combinations of $R_1$ from 3-10 au and $R_2$ from 10-25 au. f$_R1$ is the fraction of material from R$_1$, with the value of $R_1$ and $R_2$ given in the plot title. The
  contribution of R$_1$ to the total composition was varied from 5\%
  to 95\%.  Shown here are the calculated $\chi^2$ for the best fit
  $R_1$ and $R_2$ over all times and contributions. The symbols show
  where all observed molecules are matched by the models to within a
  factor of 10. }
\end{figure}

\begin{figure}
\includegraphics[width=0.45\linewidth]{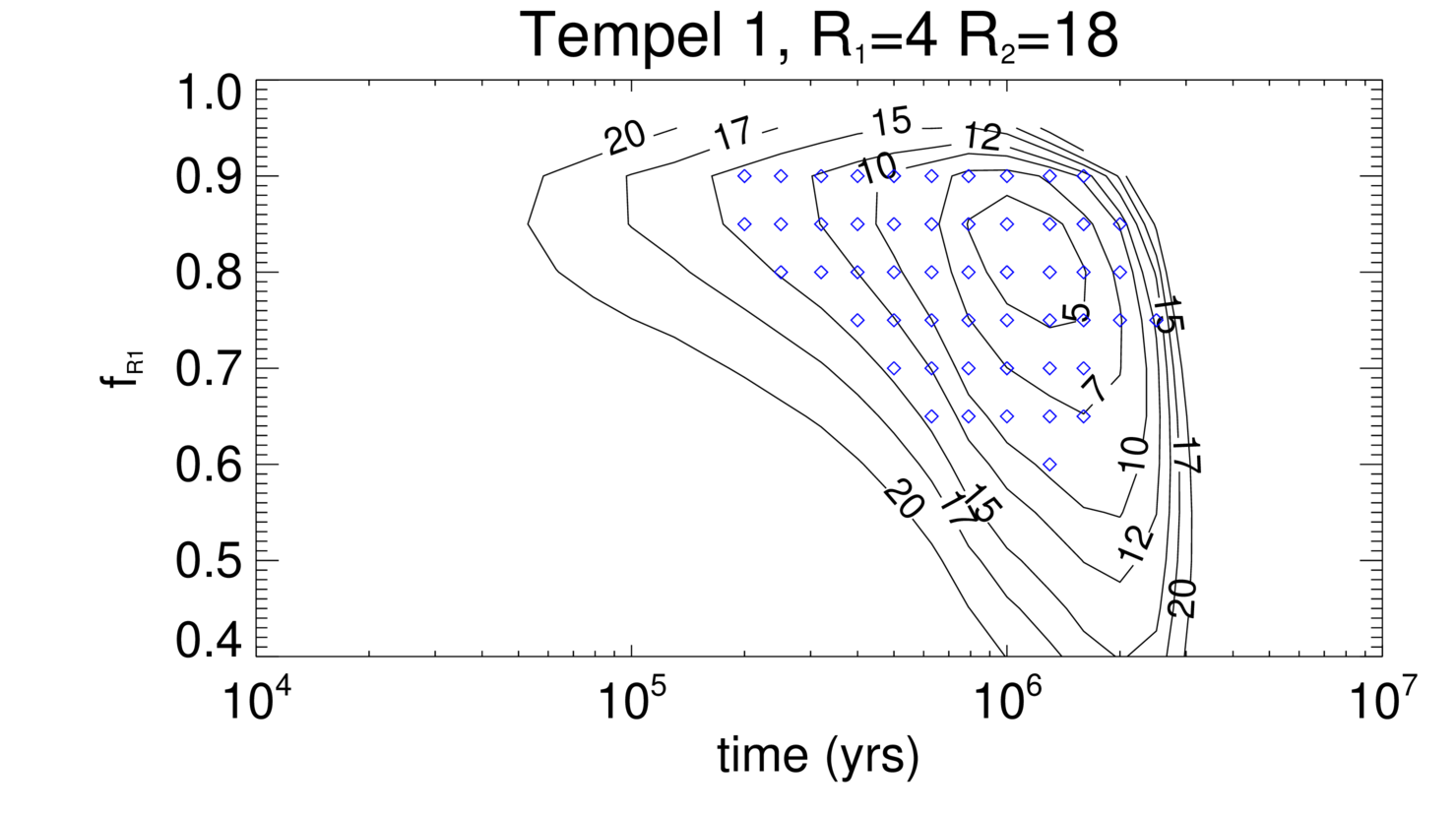}
\includegraphics[width=0.45\linewidth]{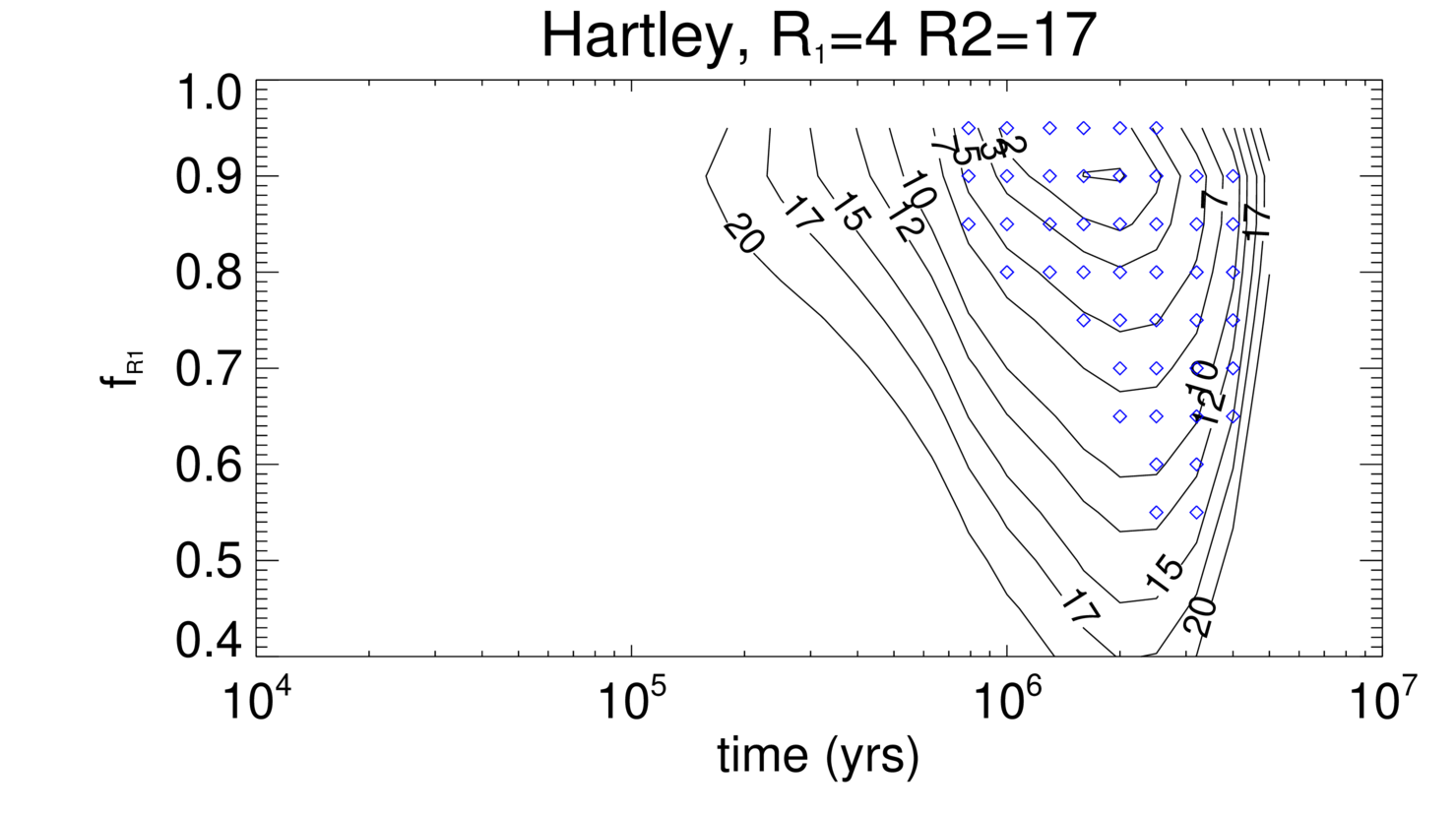}
\includegraphics[width=0.45\linewidth]{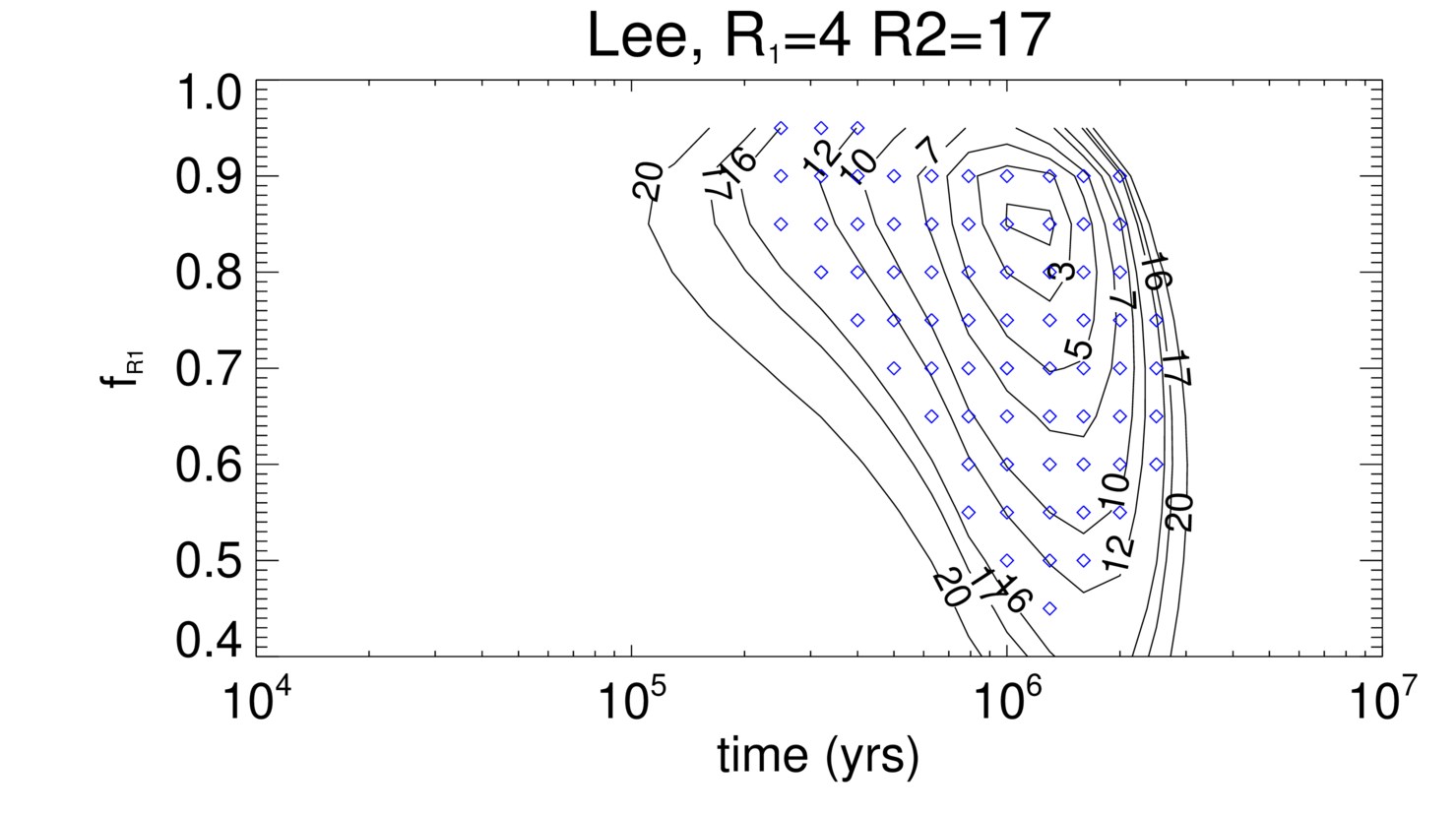}
\includegraphics[width=0.45\linewidth]{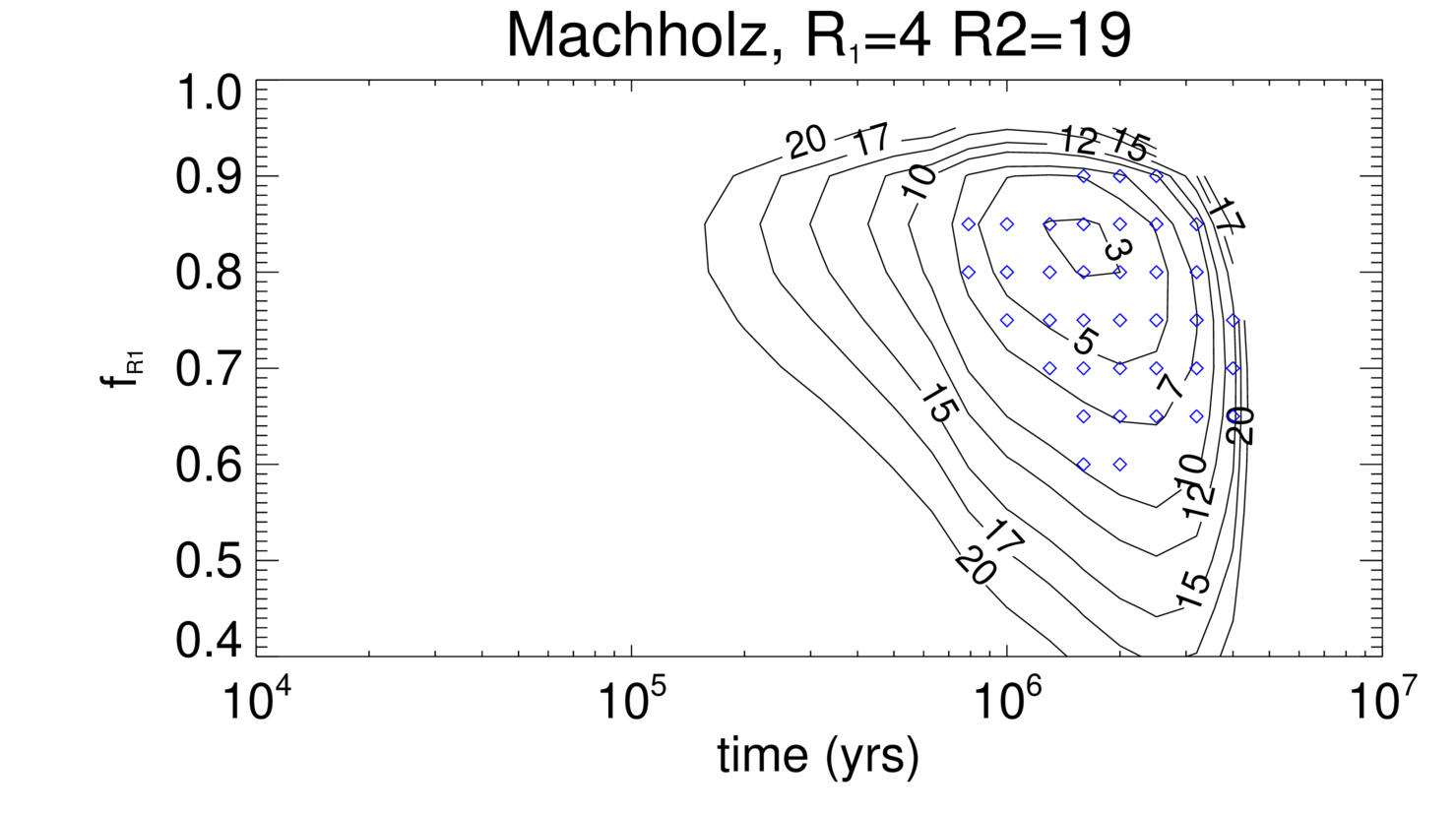}
\includegraphics[width=0.45\linewidth]{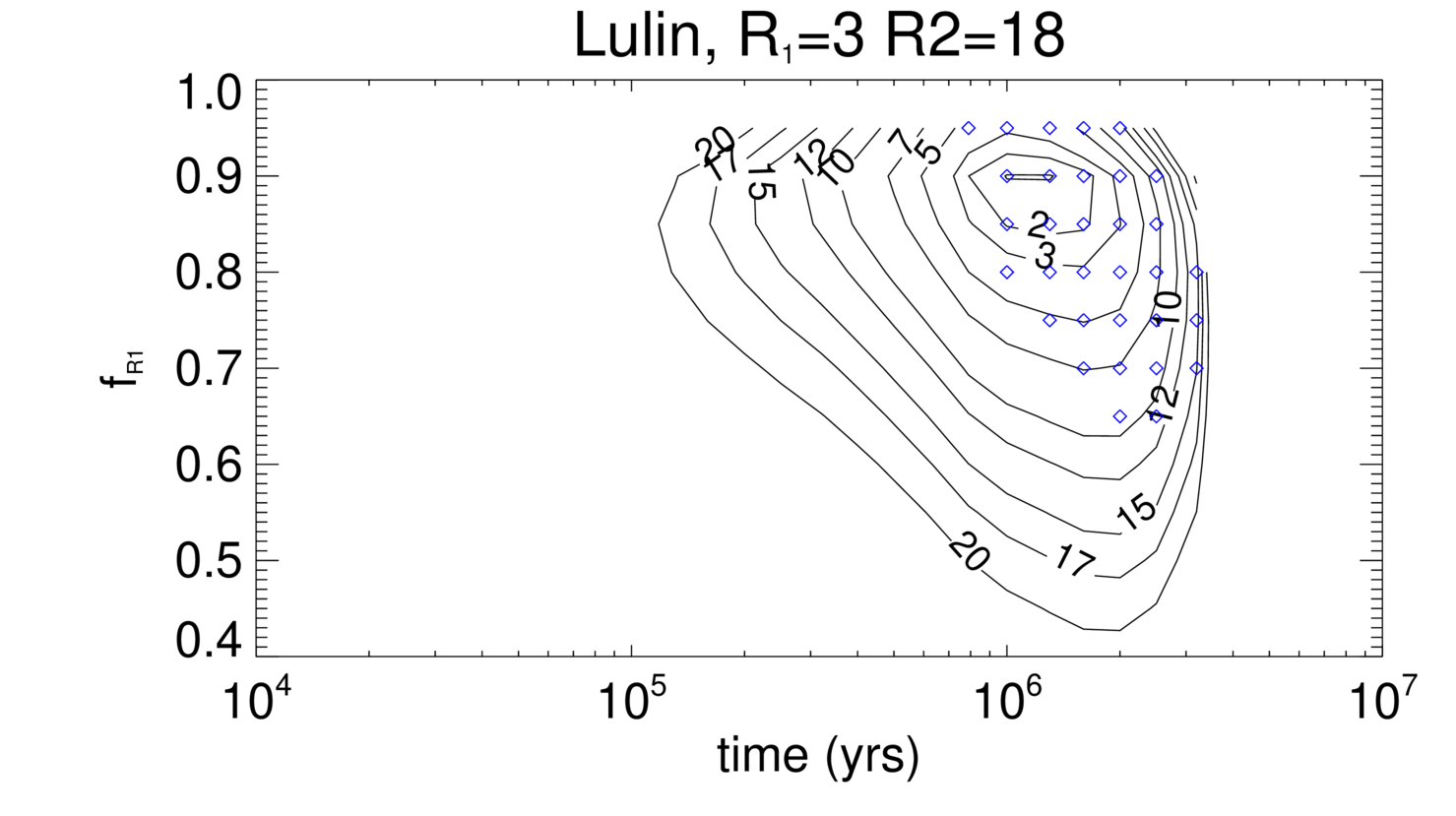}
\includegraphics[width=0.45\linewidth]{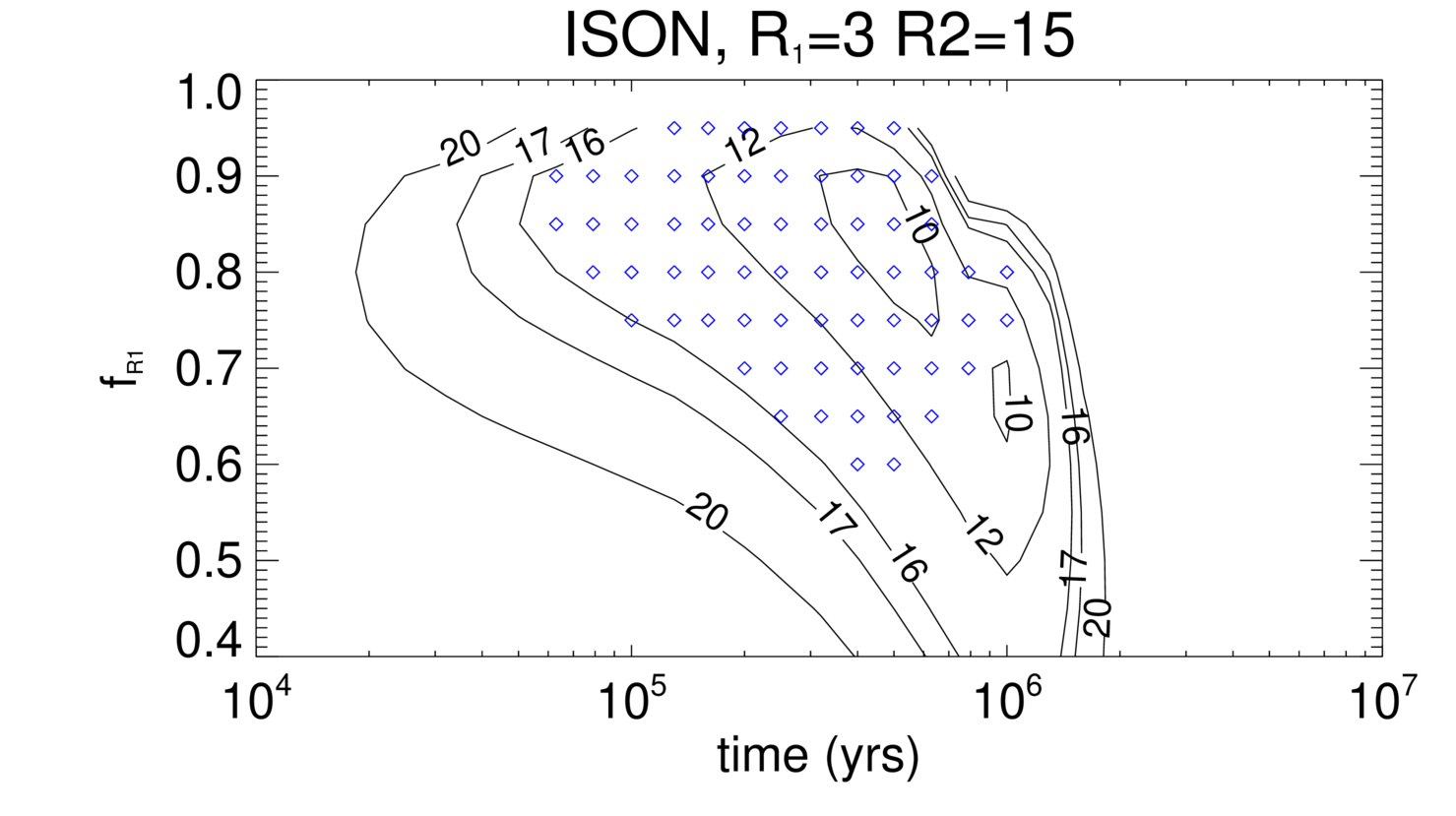}
\includegraphics[width=0.45\linewidth]{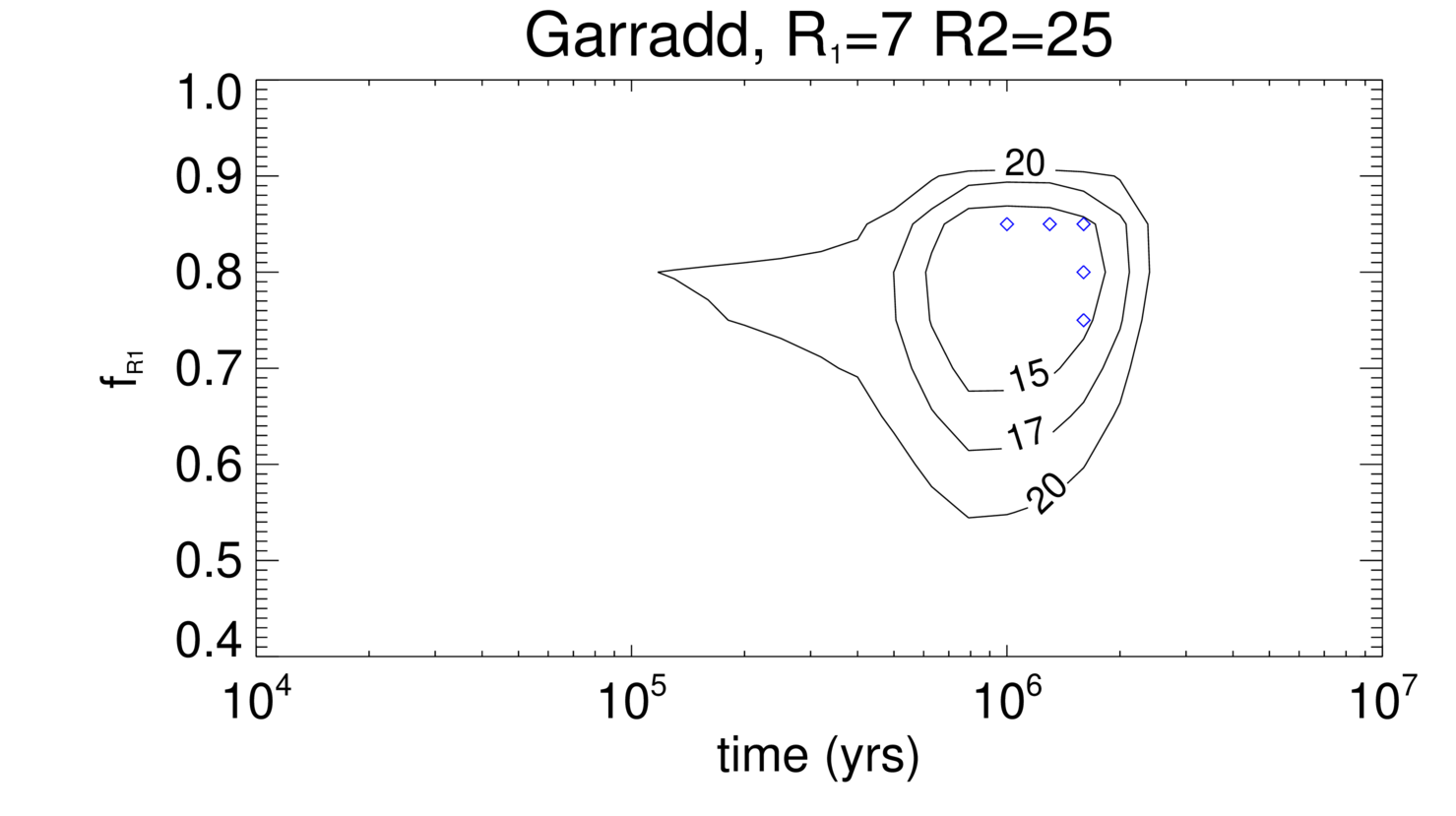}
\includegraphics[width=0.45\linewidth]{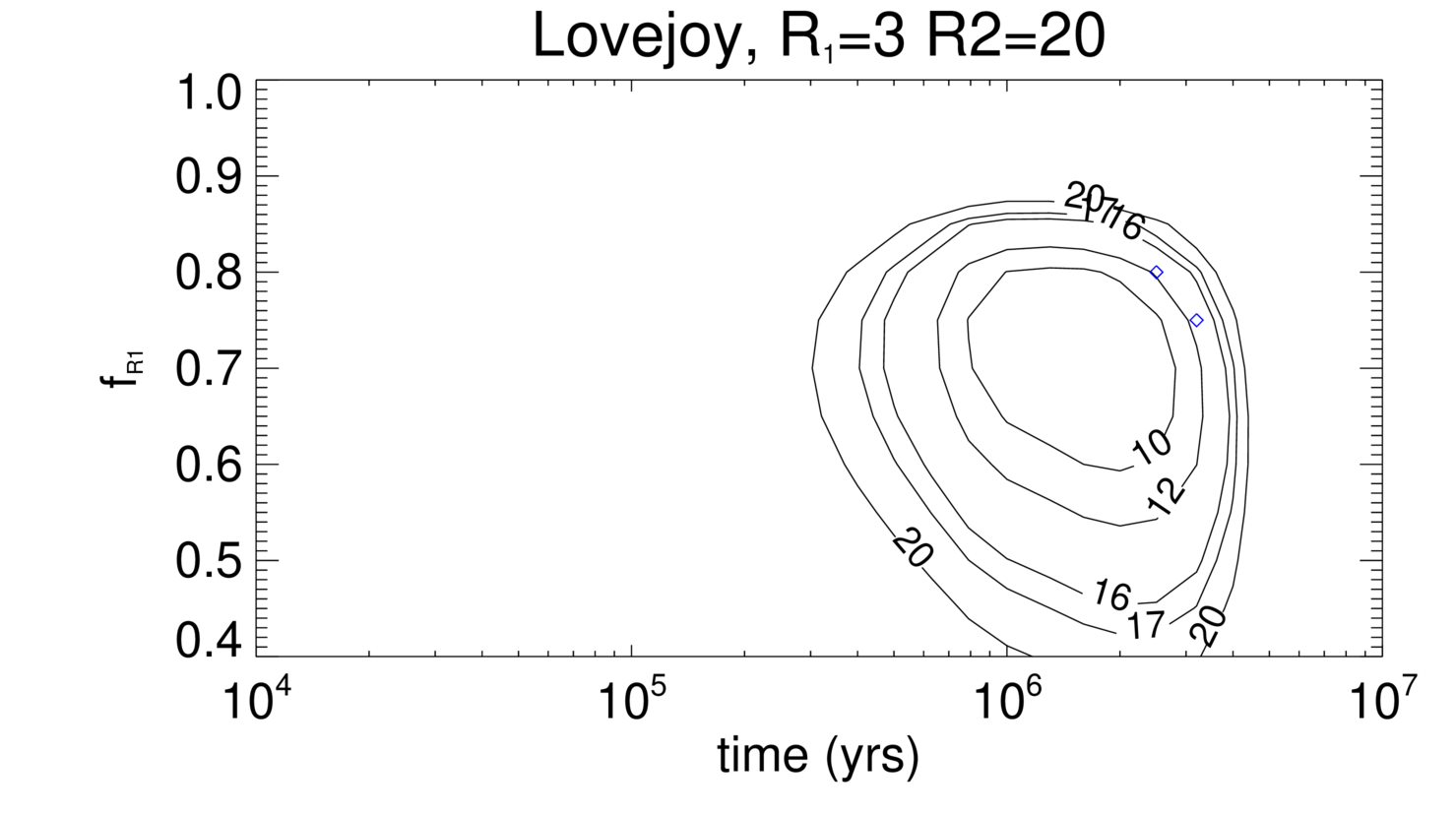}
\caption{\label{fig:2rad}The best fit models using combinations of two
  radii (one from inside the CO snowline (R$_1$), and one from outside it (R$_2$) to
  individual comets. f$_R1$ is the fraction of material from R$_1$, with the value of $R_1$ and $R_2$ given in the plot title. Contours show the $\chi^2$ values, while the blue
 diamonds indicate models where predicted mixing ratios of all
  molecules are within a factor of 10  of the observed values.}
\end{figure}

We find that for all comets considered here that there are a range of
models with different combinations of $R_1$, $R_2$ and $f_{R1}$ that
fit both  criteria.  Table~\ref{tab:2rad} lists the parameters of the
combination of models with the lowest
$\chi^2$ value.  All comets require a large contribution of material
from inside of the CO iceline.

Figure~\ref{fig:2rad_av} shows the best fit models to the average
comet data. The figure shows the contour plots of $\chi^2$ overlaid by symbols indicating 
the models for which all of the observed molecules are matched to
within a factor of 10.  For the average composition of both the OCCs
and  JFCs, the lowest $\chi^2$ is achieved
by combining models from 3 au and 18 au, but the contribution from the
two radii is different for the two families, with JFCs requiring 90\%
from 3 au compared to 80\% for OCCs.  There  are a range of
contributions from $R_1$ and $R_2$ that satisfy both criteria
for good fits (i.e. $\chi^2$ $<$ 15.5 and all abundances within an
order of magnitude of the observations). These models are
indicated on the plot by the blue diamonds.  Models fit the data
for the average OCC composition for both criteria with a contribution
from 40--90\% from 3 au at times between 0.7 and 1.2 Myr.  JFCs require
a contribution of between 70 and 95\% from 3 au at times from 0.8 to 3
Myr. This would suggest that the
JFCs formed from warmer material than the OCCs.  It should be noted
that there are other values of $R_1$ and $R_2$ that are also able to
match the observations. 

The same information for the individual comets is shown in
Figure~\ref{fig:2rad}.  One combination of radii is shown (the
one for which $\chi^2$ is lowest), but there is a range of different combinations of
$R_1$ and $R_2$ that have $\chi^2$ below the value
required for a good fit and where the model abundances are within a factor of 10 of the observations.  All of the comet compositions can be matched by
combinations of models from two radii.  It should be noted that while for most of the comets the model with the lowest $\chi^2$ also predicts abundances that are all within a factor of 10 of the observations this is not the case for Lovejoy.  For this comet the models that satisfy both fitting criteria have a slightly larger value of $\chi^2$ than the minimum.
The models here are not
sufficient to pin down the exact formation location and history of
comets but do illustrate that more than one component is required to fit
the observations.

  \begin{deluxetable}{llllll}
    \tablecaption{\label{tab:2rad}Best fit parameters assuming
      comets are made from material from two radii in the fiducial model, one inside ($R_1$) and one outside ($R_2$) the snowline. Shown are the combinations of $R_1$ and $R_2$ with the lowest $\chi^2$ which also match the observed abundances of each molecule to within a factor of 10.  The contribution of each radius is given as a percentage of the total calculated composition (so percentage of $R_1$ given is 100 x $f_{R1}$ where $f_{R1}$ is the fraction of $R_1$ in Equation~\ref{eq:frac})}.The composition of all of our comet sample can be matched by such combined models. Good fits are provided by models with $\chi^2$ $<$ 14.07 for comets with eight observations (8P/Temple 1, 103P/Hartley 2, C/1999 H1 (Lee), C/2004 Q2 (Machholz) and C/2007 N1 (Lulin)), and $<$ 15.5 for those with nine observations (C/2009 P1 (Garradd), C/2012 S1 (ISON) and C/2013 R1 (Lovejoy)). 
      \tablewidth{0pt}
      \tablehead{
 \colhead{Comet} & \colhead{time} & \colhead{$R_1$ (\%)} &
\colhead{$R_2$ (\%)} & \colhead{$\chi^2$} 
}
\startdata
9P/Tempel 1  & 1.0 Myr & 4 (85\%) & 18 (15\%) &  3.9 \\
103P/Hartley 2     & 2.0 Myr & 4 (90\%) & 17 (10\%) & 0.9  \\
C/1999 N1Lee          & 1.3 Myr & 4 (85\%) & 17 (15\%) & 1.8\\
C/2004 Q2 Machholz & 1.6 Myr & 4 (85\%) & 19 (15\%) & 2.8 \\
C/2007 N1 Lulin       &1.0 Myr & 3 (90\%) & 18 (10\%) & 0.9 \\
C/2012 S1 ISON       & 0.5 Myr & 3 (85\%) & 15 (15\%) & 9.4  \\
C/2009 P1 Garradd  & 1.6 Myr & 3 (75\%) & 19 (25\%) & 6.8 \\
C/2013 R1 Lovejoy  & 2.5 Myr & 3 (80\%) & 20 (20\%) & 12.3 \\
Average OCC & 1.0 Myr & 3 (80\%) & 18 (20\%) & 4.8 \\
Average JFC & 1.6 Myr & 3 (90\%) & 18 (10\%) & 2.0 \\
\enddata
\end{deluxetable}

\section{\label{sec:discussion}Discussion}

We have constructed a model of the chemistry of the protosolar nebula
and investigated whether this can account for the current observed
average, minimum and maximum abundances of nine molecules in comets: CO, CH$_4$,
H$_2$CO, CH$_3$OH, C$_2$H$_2$, C$_2$H$_6$, NH$_3$, HCN, and OCS.  Two
of our models -- the fiducial (with $\zeta_{CR}$ = 1.3 $\times$
10$^{-17}$ s$^{-1}$) and low cosmic ray ($\zeta_{CR}$ = 1.3 $\times$
10$^{-17}$ s$^{-1}$) -- are successful in reproducing the current
observed range of mixing ratios in comets. 
The
ices of six of the molecules we consider (CO, CH$_4$, NH$_3$, H$_2$CO, CH$_3$OH and HCN)
appear to be
inherited from the parent molecular cloud and either reflect the
molecular cloud abundances with little change (CO), or require partial
reprocessing during 
infall, or in the disk by thermal desorption, cosmic rays, or grain
surface reactions (HCN, NH$_3$, CH$_4$, H$_2$CO and CH$_3$OH).  Although the observed range of each molecule can be matched by the fiducial and low cosmic ray models, no single time and location in any of our models can match all  nine of the molecules simultaneously.

CO is not produced in our fiducial disk model, hence the highest abundance predicted is set in the molecular cloud model.  However, since CO is extremely volatile it is likely that its abundance was affected by desorption during the disk formation process and would therefore be lower than its molecular cloud value.  This would create a problem for the models to be able to account for the more CO-rich comets.  It is possible that some of the observed CO comes from the destruction of more complex molecules \citep{disanti99}, meaning that the observed abundance will be higher than that contained in the ices of the comet nucleus.

Cosmic rays play an important role in determining the abundance
distributions of NH$_3$, H$_2$CO and CH$_3$OH.  
The flux of cosmic
rays in disks is still uncertain and there is evidence from current
PPDs that the flux could be reduced (perhaps significantly) in the
inner 100 au \citep{seifert21,cleeves13}.  While reducing the cosmic
ray flux by a factor of $\sim$ 10 still allows our models to account
for the full range of observations but at later times than the
fiducial model, reducing $\zeta_{CR}$ still further results in little
change in the ice abundances compared to the molecular cloud.  A model
with such low cosmic ray flux would require photoprocessing of ices.
This could be achieved by mixing within the disk, either vertical or radial.

Other molecules, such as HCN and CH$_4$, agree best with
the cometary data near their snowlines, where their 
abundances are reduced compared to their initial values, and hence
brought closer to the cometary values.  Another way to bring their abundances into agreement with observations would be if they were formed less efficiently in the molecular cloud.  Both molecules form by hydrogenation on the grain surfaces: HCN from CN and CH$_4$ from carbon atoms. Reducing their abundance in the molecular cloud model could extend the range of radii over which the disk models can match the observations.
CH$_4$, like CO, is also very volatile, and therefore some methane ice may be expected to be lost  during the star/disk formation.

C$_2$H$_2$ and C$_2$H$_6$ differ from the molecules above in that their predicted molecular cloud values are lower than
in the comets, requiring formation in the disk, either by freezeout of
gaseous hydrocarbons or their ions, or by hydrogenation
of unsaturated hydrocarbons in the ices.

Our `reset' model, which
assumes any molecular cloud chemistry is wiped out during the disk
formation, is less successful and cannot account for the full range of observed abundances. In particular, the CO abundance, while showing a similar distribution to the fiducial model, is considerably lower. This also leads to lower CH$_3$OH and H$_2$CO abundances. Because the grains are warmer in the disk compared to the molecular cloud the formation of these two molecules is not as efficient since the lifetime of H atoms on grains is shorter.  Instead, CO is processed into CO$_2$, HNCO and OCS.

To quantify the ability of the models to account for the observations we use the $\chi^2$ test.  Based on this 
 the single location and time in the model where the ices
best match the cometary observations is near the CO snowline.
However, the values of $\chi^2$ are too high to be classed as a good fit and in addition,
the abundances of several species 
differ by more than an order of
magnitude between the model disk and the comets.  Other locations in
the model disk better fit some of these discrepant ices.  For example,
HCN only matches the observations well inside the CO snowline. Ammonia on the other hand, has a peak in abundance near the CO snowline, and the best match to the observations are away from this location. Because no
single location and time is a good match for the compositions of the
tested individual comets,
it seems plausible that these bodies incorporate materials formed at
different distances from the young Sun.  

Since the single location/time models are not able to account for the overall comet compositions we consider an alternative scenario. 
Comets comprise a large number of
molecules with very different desorption temperatures, and although
the molecules are also present in the interstellar medium their
abundances are different indicating processing is required in the
disk. The predicted abundance of  HCN in the molecular
cloud model
is considerably above the range of values seen in comets and
therefore some loss of this species needs to occur in the disk. This
only happens in regions where the temperature is too high to allow
more volatile species such as CO to remain on the grains,
suggesting that comets are made up of some combination of material that
has remained at cold temperatures and thus retains its interstellar
composition, together with other grains that have lost at least some
of their volatiles through processing at warmer temperatures.
Combinations of this kind could arise because by the transport of gas or solid bodies within the disk between locations providing distinct temperatures, pressure and radiation environments.

To explore how the observed comet compositions could have been achieved we
 constructed a toy model that combines material from inside and
outside the CO snowline. This dividing line was chosen because of the
need to retain very volatile ices in our model comet. We find that
combining material from these two regions can indeed provide a
better agreement with comet observations for all of our sample, as
well as the average comet compositions for both JFCs and OCCs.
 There is a wide range
of model radii and times that can be combined to match the
observations.  More detailed modeling is required to constrain how
particular comets obtained their particular compositions. 

There is no obvious difference between the combination of material
required for the individual comets comprising two JFCs and the seven
OCCs.  For the two-radii model
the contributions to the JFCs and OCCs come from over-lapping
regions.  This is consistent with the conclusions of \cite{ahearn12} based on
observations of the three molecules CO, CO$_2$, and H$_2$O, but is in
conflict with the classical picture of comet formation in which JFCs were
formed outside of Neptune's orbit and OCCs closer to the star.  
However, the number of comets observed in
each family is small, especially for JFCs, and therefore it is
difficult to draw definitive conclusions about the differences between
the two families based on the current observations. However, from the
average compositions it does appear that the JFCs require a higher
fraction of their composition to be CO-poor.  
To further understand the links between
comets and the protosolar nebula will require remote observations of
more comets, models treating the disk's chemistry alongside its gas
dynamics and grain growth and transport, and ultimately, close-up
visits by spacecraft to cometary nuclei with a range of dynamical
histories.

\subsection{Transport in disks}
The mix of CO-rich and CO-poor material required for the disk models to match the comet observations suggests efficient transport in the protosolar nebula.  The exact nature and efficiency of such mixing is beyond the scope of this paper, but here we briefly discuss some of main points.

Evidence for transport in the protosolar nebula comes from the presence of 
 crystalline silicates in comets  \citep[e.g.][]{brownlee09}.  Cosmic rays destroy silicates' crystalline structure in the interstellar medium \citep{kemper04}, so the comets crystalline silicates cannot be interstellar in origin and must have been heated to temperatures near 1000~K found in our young solar system only close to the Sun.  How enough crystalline material was carried out beyond the CO snowline is unclear.  Radial transport of crystalline silicates appears to occur in contemporary protostellar disks \citep{vanboekel04}.  

The motions of solids in a disk depends on their size.  Small grains can be transported with the gas and outward motion can be achieved by advection or by turbulent diffusion \citep[see reviews by][]{testi14,turner14}.  Mixing icy grains has previously been suggested as a means of accounting for the D/H ratio observed in comets \cite[e.g.][]{mousis00,yang13}. Turbulence leads to small particles making random walks through the disk, giving each its own thermal history.  Pebble-sized and larger particles partly decouple from the gas and drift towards the star near the midplane.  Outward transport would need to occur early, and primarily through the movement of small grains \citep{ha10}.

The evolution of the disk structure will also play a role in the thermal history of grains and ices.  Dust grains and their associated ices will be carried outwards by the radial expansion of the disk, potentially moving grains from warm to colder regions.   As the disk expands, material continues to fall onto it and this could retain interstellar ice signatures.  The resulting icy grains would therefore be a combination of warm material accreted earlier in the disk history with the colder, newer material. Also young stars' accretion rates generally decline with age, reducing their luminosity and thus the temperature at a given distance \citep{kh95}.

Two recent papers have considered the effects of transport in disks on ice compositions.  \cite{price21} were able to explain the exceptionally high CO/H$_2$O ratios observed in comets 2I/Borisov, C/2010 R2 (PanStarrs) and C/2009 P1 (Garradd) by inward drift of icy grains. As grains travel inwards they leave a region behind them that is depleted in water ice.  Once inside of the CO snowline the CO will desorb and the gas will be transported outwards where it can recondense in the water-ice-poor region, resulting in higher CO/H$_2$O ice abundance ratios. Similar results have been found by \cite{meijerink09} and \cite{rj13}.  

\cite{bc21} trace the evolution of ices on grains as they are transported in a disk and suggest pebbles formed at large radii drift inwards to the comet formation region with their ices preserved, so that comets form from material formed or processed at large radii. Their model includes photoprocessing of ices at $>$ 100 au, something which is not considered in our work and which can lead to conversion of simple ice molecules into more complex ones.

\begin{figure}
  \includegraphics[width=\linewidth]{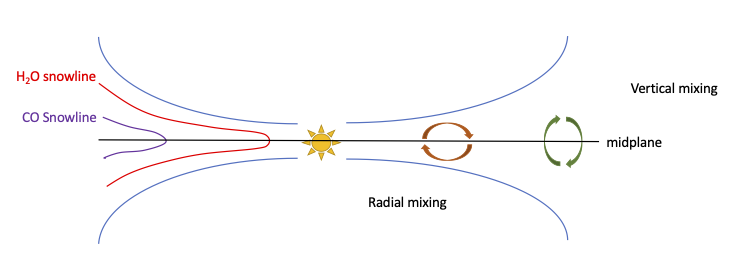}
  \caption{\label{fig:COsnow}The location where CO desorbs depends
    on both the radial and vertical location. While in the midplane
    there is a clear snowline at the radius at which the desorption
    temperature of CO is reached, there is also one above the midplane
    where the gas and dust closer to the disk surface is heated
    sufficiently.  This means that CO-depleted ice can have its origin
    either at small radii in the midplane, or above the midplane at
    larger radii. Consequently mixing from either direction (or both)
    could be the source of CO-poor material suggested by our modeling}
  \end{figure}

Here we have focused here on midplane chemistry but we do not model dynamics. Our toy model discusses combining  CO-rich and CO-poor material in the midplane which could be achieved by radial transport of grains, or by expansion of the disk and the corresponding change in physical conditions. Another plausible way of achieving this is mixing in the vertical direction. 
 Since the temperature increases with height above the midplane there is also a snowline some distance above the midplane (Figure~\ref{fig:COsnow}), and material mixed upwards will undergo  thermal processing in the same way as midplane material does as the radius decreases.  Ices that have been moved vertically will also be subject to an increased radiation field, further altering their chemistry.   
Grains fall towards the midplane where they coagulate to form larger solids, removing dust and volatiles molecules from the disks upper layers \citep{krijt16}.  However disks observed today generally show some dust in their upper layers \citep{natta07}.  Sustaining this distribution likely requires collisional fragmentation of bigger solid bodies and upward transport of the next generation dust.

The works of \cite{price21} and \cite{bc21} also provide potential mechanisms for combining CO-rich and CO-poor ices. Inflow of grains followed by evaporation of volatiles at their snowlines, and their outward transport in the gas could  enhance the abundances of molecules such as CO and CH$_4$ in ices at larger radii.   Similarly if less volatile species such as HCN behave in a similar way to H$_2$O then the abundance of these molecules could be reduced at large radii.  \cite{price21} only considered CO and H$_2$O and it is not clear how this process can affect the abundances of other species in the ices. The photoprocessing of ices in the outer disk as discussed by \cite{bc21} and the formation and inflow of pebbles from this region to the comet formation region will also  lead to changes in the ice abundance ratios.

These papers show that there is a complex interplay between chemistry and dynamics in disks that is still not fully explored. \cite{price21} only considered CO and H$_2$O and were specifically looking to account for very high CO/H$_2$O ratios.  Might this scenario also be able to account for the range of observations of all molecules in other comets?  The more volatile species such as CH$_4$ might be expected to follow the behavior of CO and be enhanced in some regions of the disk as a result of transport in the gas from inside of their snowline to regions where the water ice is depleted.  And less volatile molecules such as HCN may follow the behavior of water, being transported on grains inwards and leaving behind a region where their abundance is reduced.  The inclusion of outer disk photoprocessing of grains by \cite{bc21} could also provide a way to generate the observed range of molecular abundance ratios, with formation of more complex molecules out of simpler ones changing the abundances.

The important point here is that the models suggest that a combination of  CO-rich (cold) and CO-poor (warm) ices is required to account for comet compositions.  How this occurs is still unclear - the interplay of chemistry and transport is clearly very complex and still not well understood.  In practice it is likely that many processes affect the composition of cometary ices, including radial and vertical transport of small grains, as well as inward movement of larger pebbles.  Further more detailed model is required to assess how these processes led to the variety of comet compositions seen in the solar system.

\subsection{Comparison to the Eistrup Model}
The $\chi^2$ tests in Section~\ref{sec:1rad} suggest that the best
way to fit the comet observations by a model at a single location and time is for them to be assembled near the
CO snowline. 
Comet compositions were also linked to the CO snowline by
\cite{eistrup19}, who used a maximum-likelihood function to determine
where their model disks best fit measurements of individual comets.
Their best fits roughly track the location of the CO snowline, but they found that the `reset' model was a better fit to the observations than the `inheritance' model (the E19 `inheritance' model is equivalent to our fiducial model).  

While an exact comparison between the two models and their reaction networks has not been made the differences are likely to originate in two factors. First, our model does not include the evolution of the disk and changes in the radial distribution of density and temperature, and the corresponding movement in the location of the icelines is likely to have an effect on the chemistry.   
The E19 CO snowline moved inwards with time,
from about 28~au at 1~Myr to 12~au at 8~Myr.   This cooling will
affect the abundance of the more volatile species such as CO and N$_2$
because they will condense at progressively smaller radii as the disk
evolves. 
  
  Second, there are differences in the grain chemistry.  Both E19 and our current work use a two-phase rate equation model but there are differences in the way that the grain chemistry is treated. For example, a key parameter is the ratio of the barrier to diffusion ($E_D$) to the binding energy ($E_b$).  E19 assume a ratio of 0.3, whereas we use 0.34.  Both values are within the range suggested by the Monte Carlo simulations of \cite{kc14}.  Our value was chosen to provide a reasonable match between our models and the observations of ices in molecular clouds (see Table~\ref{tab:ice}).  The slightly higher value in this work will slow down the rate at which grain reactions can occur.  Another difference is in the number of active layers assumed.  Here we choose 4, whereas E19 have 2. Again, this will make a difference to the reaction rates.    There may also be differences in the grain reactions included and activation barriers assumed in the two models.

These differences in the model assumptions result in differences in the predicted mixing ratios. Comparing our Figure~\ref{fig:reset} with the `reset' model in Appendix A of E19 we can see both similarities and differences. 
The maximum CO/H$_2$O ratio in the E19 model is 1\%, similar to the value we find over much of the disk, and well below the highest observed values of $>$ 20\% in their comet sample.  In their model the CO appears to have been converted into H$_2$CO and CH$_3$OH, whereas these molecules are under-produced in the current work, with CO being processed into CO$_2$, HNCO and OCS.   These differences are likely due to differences in the grain chemical networks.  

Both E19's `reset' and `inheritance'  models have similar CH$_4$ distributions which in turn are similar to ours.  The best matches to the observed abundances for all of the models are around the CH$_4$ snowline.  Outside of this the abundance relative to water is higher than observed. 

For the hydrocarbons, E19's `reset' and `inheritance'  models produce quite different distributions.  C$_2$H$_2$ is only produced in the `reset' model between 5 and 10 au and then only at high enough abundances to match the observations at times earlier than 2 Myrs.  The C$_2$H$_2$ in our `reset' model also peaks between 5-10 au like E19's but with smaller mixing ratios. 

C$_2$H$_6$ is more widespread than C$_2$H$_2$ in both of E19s models and can be $>$ 10\% relative to water in some parts of the disk.  Their `reset'  model provides a match to the range of observations on either side of the CO snowline, as does ours.

The observed HCN ratio in comets are $<$ 0.5\% relative to H$_2$O.  In our fiducial model this only occurs near the HCN snowline, but can be achieved at larger radii in our `reset' model.  The E19 `inheritance' model finds similar mixing ratios (HCN/H$_2$O $\sim$ 1\% between 5-10 au (a wider radius range than in our model) and also outside of the iceline, where our abundance is much higher.  However, for their `reset' model HCN/H$_2$O is greater than 10\% everywhere in the disk, much higher than the observed value. 

E19's `reset' model produces considerably more OCS than their `inheritance' model.  Both models cover the observed range of abundances in the region of the CO iceline.  Our fiducial model differs, in that the OCS abundances are best fit by models outside of 20 au and at late times ($>$ 10 Myrs), or near the OCS iceline. In our `reset' model OCS only matches the observations near its snowline.  For the rest of the parameter space our model over produces  OCS.   

In summary, the differences between our model and that of E19 are likely to lie in differences in the details of the grain chemistry.

\section{\label{sec:concl}Conclusions}
Our conclusions can be summarized as follows:
\begin{enumerate}
\item The range of comet compositions can only be matched if interstellar ices are retained.  Starting the disk chemistry with `reset' abundances does not provide a way to account for the composition of the observed comets.  Note: a hybrid model where some, but not all, of the interstellar ices are retained could provide a means of accounting for the observations but is not explored here.
\item The location and times at which models can match the range of
  mixing ratios observed for individual molecules do not overlap. No single place or time can  account for the
  overall composition of any individual comet, nor for the average composition of the comet families.  
\item  Transport has a major role to play in explaining comet volatile compositions.  Warm (CO-poor) and cold (CO-rich) material must be combined to account for the mean abundances of the observed  Jupiter family and Oort cloud comets. The combination could be made by either radial or vertical transport along the disks' temperature gradients.
\item Comparison of the models to the average Jupiter family and Oort cloud comet compositions suggests that the the families formed in (at least) partly overlapping regions of the disk. Thus, independent conclusion from disk chemistry (this work) and the dynamical models of the early Solar System (based on the Nice model) are consistent. However, the observations to date, suggest that Jupiter-family comets contain less CO than Oort cloud comets. If this difference is related to formation processes our models suggest that Jupiter-family group comprises more warmer material than Oort cloud comets. More definitive interpretation is pending an increased sample of CO detections in Jupiter-family comets.
\item Abundances of individual comets may be more affected by post-formative processes than the ensemble properties (compositional ranges and average).  Nevertheless, a combination of warm (CO-poor) and cold (CO-rich) material also accounts for the observed composition of eight individual comets.
\end{enumerate}

None of the combined models presented here should be taken as exact
descriptions of where or when the comets formed.  Rather, they show
that ices from different places in the protosolar nebula are required to
produce compositions similar to those of comets.  A single location
and time is not sufficient to explain the abundances. The next logical
steps in modeling involve understanding the effect of the mixing on
the evolution of ice chemistry in PPDs.

Comets assembled out of materials processed for different lengths of
time or in different regions of the disk would complicate constraining
the location and epoch of formation solely through remote
observations. If comets were assembled over a period of time, then
solid particles' history of growth and transport would have played a
key role in determining the composition of the ices incorporated.  As
grains coagulated into bigger bodies, the ices deep inside would have
become shielded from photons and cosmic rays and thus less likely to
suffer alteration.  These issues could be explored using models
combining the chemistry with the transport of gas and dust and the
growth and fragmentation of solid particles. If comet nuclei contain chemically-distinct
components at the scale of boulders, pebbles, or grains, then
determining their individual compositions to infer the conditions
under which the components formed might require in-situ measurements
or even sample return.

%%%%%%%%%%%%%%%%%%%%%%%%%%%%%%%%%%%%%%%%%%%%%%%%%%%%%%
\begin{acknowledgments} The disk modeling research presented in this paper 
was conducted at the Jet Propulsion
Laboratory, California Institute of Technology, under contract with
the National Aeronautics and Space Administration and with the support
of NASA's Emerging Worlds Program award number 15-EW15-2-0150 (Willacy, Turner).  
Integration with comet observations was also supported by NASA EW80NSSC20K0341 (Dello Russo, Vervack), and NSF AST-2009398 (Bonev) and AST-2009910 (Gibb). The
project developed out of workshops on `Exploring the Early Solar
System by Connecting Comet Composition and Protoplanetary Disk
Models,' held in 2016 and 2018 at the International Space Science
Institute, Bern.  
Some kinetic data used in this paper have been downloaded from the online
database KIDA (Wakelam et al. 2012, http://kida.obs.u-bordeaux1.fr). 
© 2022. All rights reserved.
\end{acknowledgments}
\clearpage
\newpage
\appendix

\clearpage
\section{\label{app:kida}Reactions included from the kida database}. 

\startlongtable
\begin{deluxetable}{lccr}
  \tablecaption{\label{tab:kida}Neutral-neutral reactions added to
    the UMIST RATE12 database \citep{rate12} 
from the KIDA database \citep{kida}.  Rates are given by the Arrhenius
    equation \mbox{$k_j = f_j (T/300.)^{\alpha_j} exp(-\beta_j/T)$
    cm$^{-3}$ s$^{-1}$.}}
  \tablewidth{0pt}
  \tablehead{
    \colhead{Reaction} & \colhead{f$_j$} & \colhead{$\alpha_j$} &
    \colhead{$\beta_j$}
  }
  \startdata
  O +  CCS = CO + CS                                             &       1.00 $\times$ 10$^{-10}$  &   \\
H       +     CH$_3$CH$_3$   =    H$_2$      +     C$_2$H$_5$ & 1.22 $\times$ 10$^{-11}$   &   1.5  &    3720.0      \\
CH$_2$     +     CH$_3$CH$_3$  =     CH$_3$      +    C$_2$H$_5$ &                     1.07 $\times$ 10$^{-11}$   &  0.0 &   3980.0   \\
CH$_2$     +     CH$_3$CH$_3$   =    CH$_4$       +   C$_2$H$_4$   &                   1.82 $\times$ 10$^{-16}$   &  6.0  &  3040.0     \\
CH$_3$      +    CH$_3$CCH    =   C$_2$H        +  CH$_3$CH$_3$&                    8.32 $\times$ 10$^{-13}$   &  0.0 &   4430.0     \\
 C$_2$H     +     CH$_3$CH$_3$    =   C$_2$H$_2$    +     C$_2$H$_5$    &                  5.10 $\times$ 10$^{-11}$   &  0.0  &    76.0    \\
C$_2$H$_3$    +     C$_2$H$_5$     =    C$_2$H$_2$     +    CH$_3$CH$_3$    &                9.80 $\times$ 10$^{-12}$  &    \\
C$_2$H$_3$   +      CH$_3$CH$_3$   =    C$_2$H$_4$ +        C$_2$H$_5$        &              1.49 $\times$ 10$^{-13}$ &    3.3  &  5280.0  \\
H$_2$     +      C$_2$H$_5$     =    H         +   CH$_3$CH$_3$         &           4.23 $\times$ 10$^{-15}$    & 3.6  &  4250.0     \\
CH$_4$     +     C$_2$H$_5$     =    CH$_3$       +   CH$_3$CH$_3$        &            2.57 $\times$ 10$^{-15}$    & 4.14 &   6320.0     \\
 C$_2$H$_2$   +      C$_2$H$_5$    =     C$_2$H      +    CH$_3$CH$_3$       &             4.50 $\times$ 10$^{-13}$   &  0.0 &  11800.0    \\
C$_2$H$_4$    +     C$_2$H$_5$      =   C$_2$H$_3$     +    CH$_3$CH$_3$       &             5.67 $\times$ 10$^{-14}$   &  3.13  &  9060.0     \\
C$_2$H$_5$     +    C$_2$H$_5$      =   C$_2$H$_4$     +    CH$_3$CH$_3$        &            2.40 $\times$ 10$^{-12}$    &    \\
H$_2$CCC     +   CH$_3$CH$_3$   =    CH$_2$CCH    +   C$_2$H$_5$       &               1.90 $\times$ 10$^{-10}$   &    \\
 CH$_2$CCH   +    CH$_3$CH$_3$   =    CH$_3$CCH +      C$_2$H$_5$      &                5.83 $\times$ 10$^{-14}$  &   3.3  &  9990.0  \\
CH$_2$CCH    +   CH$_3$CH$_3$    =   C$_2$H$_5$   +      CH$_2$CCH$_2$    &               5.83 $\times$ 10$^{-14}$ &    3.3 &   9990.0    \\
 NH         +  C$_2$H$_5$      =   N     +       CH$_3$CH$_3$              &      4.00 $\times$ 10$^{-11}$  &    \\
 H +  C$_2$H$_5$ = CH$_3$ +  CH$_3$                     &            1.25 $\times$ 10$^{-10}$   &  \\
 H +  C$_2$H$_5$ = H$_2$ +  C$_2$H$_4$                &        3.00 $\times$ 10$^{-12}$  &      \\
  H + CH$_2$CCH$_2$ = H +  CH$_3$CCH                  &             1.29 $\times$ 10$^{-11}$    & 0.0 &    1160.0    \\
 H +  CH$_3$CHCH$_2$ =  CH$_3$ + C$_2$H$_4$        &               1.20 $\times$ 10$^{-11}$ &    0.0 &     655.0  \\
  CH$_2$ + CH$_3$ = H + C$_2$H$_4$                        &    7.00 $\times$ 10$^{-11}$   &   \\
   CH$_2$ + C$_2$H$_3$ = C$_2$H$_2$ +  CH$_3$        &            3.00 $\times$ 10$^{-11}$   &    \\
   CH$_2$ +  C$_2$H$_5$ = CH$_3$ + C$_2$H$_4$        &         3.00 $\times$ 10$^{-11}$  &    \\
    CH$_3$ +  C$_2$H$_4$ = C$_2$H$_3$ +  CH$_4$       &          1.61 $\times$ 10$^{-14}$  &   3.7   & 4780.0      \\
    CH$_3$ + C$_2$H$_5$ = CH$_4$ + C$_2$H$_4$         &     1.90 $\times$ 10$^{-12}$   &     \\
    CH$_3$ +  CH$_3$CH$_3$ = CH$_4$ + C$_2$H$_5$    &         1.82 $\times$ 10$^{-16}$   &  6.0   & 3040.0      \\
    CH$_3$ +  CH$_2$CCH$_2$ = C$_2$H$_2$ +  C$_2$H$_5$       &                3.32 $\times$ 10$^{-13}$    & 0.0 &    4080.0      \\
    CCH + CH$_4$ = C$_2$H$_2$ +  CH$_3$                    &    1.20 $\times$ 10$^{-11}$   &  0.0 &     491.0      \\
    CCH + CH$_3$CH$_3$ =  C$_2$H$_2$ + C$_2$H$_5$  &           5.10 $\times$ 10$^{-11}$  &   0.0 &      76.0      \\
    H$_2$ +  C$_2$H$_3$ =  H + C$_2$H$_4$                   &    3.01 $\times$ 10$^{-20}$  &   \\
    C$_2$H$_3$ + CH$_4$ = CH$_3$ + C$_2$H$_4$            &         2.18 $\times$ 10$^{-14}$  &   4.02  &  2750.0      \\
    C$_2$H$_3$ + C$_2$H$_3$ = C$_2$H$_2$ +  C$_2$H$_4$        &               3.50 $\times$ 10$^{-11}$   &     \\
    H$_2$ +  H$_2$CCC = H +  CH$_2$CCH                          &
    1.20 $\times$ 10$^{-10}$   &     \\
      CH$_4$ +  H$_2$CCC = CH$_3$ + CH$_2$CCH               &             5.90 $\times$ 10$^{-11}$  &      \\
    C$_2$H$_3$ +  H$_2$CCC = C$_2$H$_2$ +  CH$_2$CCH   &              3.00 $\times$ 10$^{-11}$ &        \\
   C$_2$H$_5$ + H$_2$CCC = C$_2$H$_4$ +  CH$_2$CCH     &          1.50 $\times$ 10$^{-11}$   &      \\
    CH$_4$ +  CH$_2$CCH =  CH$_3$ + CH$_3$CCH             &        1.74 $\times$ 10$^{-14}$  &   3.4  & 11700.0      \\
   CH$_4$ + CH$_2$CCH =  CH$_3$ +  CH$_2$CCH$_2$        &      1.74 $\times$ 10$^{-14}$   &  3.0 &  11700.0      \\
  N +  CH$_3$CN = H +  HCN + HCN                                   &       2.28 $\times$ 10$^{-15}$   &  0.0 &     813.0      \\
   NH + CH$_3$ = N +  CH$_4$                                      &     4.00 $\times$ 10$^{-11}$   &      \\
   NH + C$_2$H$_3$ = N + C$_2$H$_4$                         &     4.00 $\times$ 10$^{-11}$  &        \\
  NH$_2$ +  C$_2$H$_4$ = NH$_3$ +  C$_2$H$_3$            &           3.42 $\times$ 10$^{-14}$  &   0.0 &    1320.0      \\
   NH$_2$ +  C$_2$H$_5$ = NH$_3$ +  C$_2$H$_4$             &          4.15 $\times$ 10$^{-11}$   &     \\
   NH$_2$ +  CH$_3$CH$_3$ =  NH$_3$ + C$_2$H$_5$          &             6.14 $\times$ 10$^{-13}$    & 0.0 &    3600.0      \\
   H +  CH$_3$CN = HCN + CH$_3$                   &     3.39 $\times$ 10$^{-12}$  &   0.0 &    3950.0      \\
   H +  CH$_3$CN = CN +  CH$_4$                     &   1.66 $\times$ 10$^{-13}$   &  0.0 &   1500.0      \\
   O +  CH$_3$CH$_3$ = OH +  C$_2$H$_5$                 &      8.63 $\times$ 10$^{-12}$  &   1.50  &  2920.0      \\
    OH +  C$_2$H$_4$ = H$_2$O +  C$_2$H$_3$                 &      1.60 $\times$ 10$^{-13}$   &  2.74    &2100.0      \\
    OH +  CH$_3$CH$_3$ =  H$_2$O + C$_2$H$_5$             &          6.90 $\times$ 10$^{-12}$   &  0.0 &    1010.0      \\
    CH$_2$ +  H$_2$O = OH +  CH$_3$                      &  1.60 $\times$ 10$^{-16}$  &  \\
    H$_2$ + HCO = H + H$_2$CO                    &   2.70 $\times$ 10$^{-13}$  &   2.0  &  8980.0      \\
    HCO +  CH$_4$ = CH$_3$ + H$_2$CO             &          1.39 $\times$ 10$^{-13}$   &  2.85 &   11300.0      \\
    HCO +  C$_2$H$_3$ = CO + C$_2$H$_4$              &         1.50 $\times$ 10$^{-10}$  &     \\
    HCO +  C$_2$H$_5$ = CO + CH$_3$CH$_3$         &            2.01 $\times$ 10$^{-10}$   &     \\
    HCO +  CH$_3$CH$_3$ =  H$_2$CO + C$_2$H$_5$    &                   4.26 $\times$ 10$^{-13}$    & 2.72  &  9280.0      \\
    H$_2$O + HCO  = OH + H$_2$CO                 &      8.61 $\times$ 10$^{-13}$    & 1.35    &  13.1      \\
    HCN + HCO  = CN + H$_2$CO                 &      1.00 $\times$ 10$^{-11}$  &    0.0 &   17200.0      \\
    H$_2$CO +  C$_2$H$_3$ =  HCO + C$_2$H$_4$          &             8.22 $\times$ 10$^{-14}$   &  2.81  &  2950.0      \\
    H$_2$CO +  C$_2$H$_5$ =  HCO +  CH$_3$CH$_3$    &    8.31 $\times$ 10$^{-14}$ &    2.81 &    2950.0      \\
    C +  CO$_2$ = CO +  CO                      &   1.00 $\times$ 10$^{-15}$    &      \\
    CCH + CCH = C$_2$ + C$_2$H$_2$ & 3.00 $\times$ 10$^{-12}$ & & \\
    CCH + C$_2$H$_3$ = C$_2$H$_2$ + C$_2$H$_2$ & 1.600 $\times$ 10$^{-12}$ &  \\
    C$_2$H + C$_2$H$_5$ = C$_2$H$_2$ + C$_2$H$_4$ & 3.00 $\times$ 10$^{-12}$ & & \\
    CCH + C$_2$H$_5$ = CH$_3$ + CH$_2$CCH & 3.00 $\times$ 10$^{-11}$ & & \\
    NH$_2$ + C$_2$H$_2$ = C$_2$H + NH$_3$ & 1.11 $\times$ 10$^{-13}$ & & \\
    CH$_4$ + C$_3$H = CH$_3$ + H$_2$CCC & 1.2 $\times$ 10$^{-11}$ & & \\
    OH + C$_2$H = O + C$_2$H$_2$ & 3.00 $\times$ 10$^{-11}$ & & \\
    OH + C$_2$H = CO + CH$_2$ & 3.01 $\times$ 10$^{-11}$ & & \\
    C$_2$H + HCO =  CO + C$_2$H$_2$ & 1.00 $\times$ 10$^{-10}$ & & \\
C   +   H$_2$S     =     H       +     HCS                   &    2.50 $\times$ 10$^{-10}$   &   \\
C   +   H$_3$S$^+$    =     H$_2$S     +     C        +    H    &        7.51 $\times$ 10$^{-8}$  &  -0.50   &    \\
C        +    H$_2$CS     =    CS       +    CH$_2$                   &    1.00 $\times$ 10$^{-10}$      &      \\
H$_3$CS$^+$ +       e-      =     H$_2$S   +       C       +     H     &       7.51 $\times$ 10$^{-8}$ &   -0.50 &        \\
S         +   C$_3$H         = C$_2$H     +     CS                    &    7.00 $\times$ 10$^{-11}$     \\
S         +   C$_3$H      =    C$_3$S      +    H                      &   3.00 $\times$ 10$^{-11}$      \\
H         +   HSO$^+$   =      SO$^+$   +       H$_2$                 &       2.00 $\times$ 10$^{-10}$  &     \\
H$_2$       +    C$_3$S$^+$   =      H       +     HC$_3$S$^+$             & 4.30 $\times$ 10$^{-10}$ &       \\
C$^+$      +     C$_3$S      =    C$_3^+$    +      CS                  &      5.00 $\times$ 10$^{-10}$  &  -0.50       \\
C$^+$      +     C$_3$S     =     C$_3$      +     CS$^+$                 &      5.00 $\times$ 10$^{-10}$&    -0.50   &    \\
C        +    NS$^+$     =     CN      +     S$^+$                   &     6.00 $\times$ 10$^{-10}$  &       \\
C       +     SO$^+$     =     CO     +      S$^+$                  &      6.00 $\times$ 10$^{-10}$  &       \\
C       +     HCS     =     H        +    C$_2$S                  &     2.00 $\times$ 10$^{-10}$  &      \\
C       +     HCS     =     S         +   C$_2$H                  &     1.00 $\times$ 10$^{-10}$  &      \\
C         +   C$_2$S     =     C$_2$       +    CS                   &     2.00 $\times$ 10$^{-10}$   &      \\
C          +  C$_3$S     =     C$_3$       +    CS                    &    3.00 $\times$ 10$^{-10}$  &      \\
CH        +   CS      =     C$_2$S     +     H                    &     1.50 $\times$ 10$^{-10}$  &      \\
CH        +   CS      =     C$_2$H     +     S                     &    5.00 $\times$ 10$^{-11}$ &       \\
CH       +    C$_3$S     =     CS       +    C$_3$H                  & 1.00 $\times$ 10$^{-10}$ &      \\
CH$_3^+$   +      CS      =     CH$_3$CS$^+$ +       PHOTON      &
1.00 $\times$ 10$^{-13}$   & -1.00 & 0.0     \\
N        +    NS$^+$      =    N$_2$       +    S$^+$                  &      6.00 $\times$ 10$^{-10}$   &     \\
N        +    C$_2$S      =    CN       +    CS                    &    3.00 $\times$ 10$^{-11}$   &  0.17       \\
NH      +     CS      =     HNC      +    S                     &    1.00 $\times$ 10$^{-11}$   &  0.0   & 1200.0     \\
NH      +     C$_2$S    =      HCN     +     CS                  &      2.00 $\times$ 10$^{-11}$   &    \\
NH      +     C$_2$S    =      HNC     +     CS                  &      2.00 $\times$ 10$^{-11}$   &   \\
O        +    C$_3$S     =     CO        +   C$_2$S                  &     1.94 $\times$ 10$^{-11}$    & 0.0   &  231.0     \\
S         +   C$_2$H$_3$    =     CH$_3$      +    CS                  &      4.00 $\times$ 10$^{-11}$   &    \\
S         +   C$_2$H$_3$     =    HS        +   C$_2$H$_2$               &       1.00 $\times$ 10$^{-11}$   &     \\ 
S         +   C$_3$H$_2$    =     C$_2$H$_2$    +     CS                 &       1.00 $\times$ 10$^{-10}$  &     \\  
S         +   C$_2$S      =    CS         +  CS                    &    1.00 $\times$ 10$^{-10}$    &   \\ 
 H$_2$S$^+$     +    HNC   =       HCNH$^+$  +      HS           &             8.97 $\times$ 10$^{-10}$   & -0.50   &   \\ 
H$_2$S$^+$      +   HCN    =      HCNH$^+$ +       HS             &           9.46 $\times$ 10$^{-10}$  &  -0.50   &    \\
HSO$^+$      +   CO     =      HCO+   +      SO              &          1.00 $\times$ 10$^{-9}$    &  \\ 
HSO$^+$     +    H$_2$O    =      SO      +     H$_3$O$^+$           &           2.00 $\times$ 10$^{-9}$  &   \\
  \enddata
\end{deluxetable}

\section{\label{app:mr}Comet observations}

\begin{deluxetable}{llrrrrrrrrrr}
  \tablecaption{\label{tab:mr_ref}Observational data used in this paper. Abundances are 
    given as percentages with respect to H$_2$O.  Abundances
  in parentheses were not used by DR16 to determine the average
  comet compositions given in Table~\ref{tab:av}.  DR16 also
  discuss 17P/Holmes and C/2010 G2 but these are not included in this
  table because they were not used to calculate the average molecular 
 mixing ratios. Also shown are the
  OCS abundances from \cite{saki20}.}
  \tablewidth{0pt}
  \tablehead{
    \colhead{Comet} & \colhead{Family} & \colhead{CH$_3$OH}  &
\colhead{HCN} & \colhead{NH$_3$} & \colhead{H$_2$CO} &
\colhead{C$_2$H$_2$} & \colhead{C$_2$H$_6$} & \colhead{CH$_4$} &
\colhead{CO}  & \colhead{OCS}\\
}
\startdata
2P/Encke & JFC & 3.48 & 0.09 & \nodata & $<$ 0.13 & $<$ 0.08 & 0.31 &
0.34 & $<$ 1.77 & 0.06 \\
6P/d'Arrest & JFC & 2.8 & 0.03 & 0.52 & 0.36 & $<$ 0.05 & 0.29 & \nodata &
\nodata & \nodata\\
9P/Tempel 1 & JFC & 1.4 & 0.2 & 0.9 & 0.84 & 0.13 & 0.29 & 0.54 & 4.3
& \nodata\\
10P/Tempel 2 & JFC & 1.58 & 0.13 & 0.83 & $<$ 0.11 & $<$ 0.07 & 0.39 &
\nodata & \nodata & \nodata\\
21P/G-Z & JFC & 1.22 & $<$ 0.27 & \nodata & ($<$ 0.8) & $<$ 0.42 & 0.12
& \nodata &  2.2 & 0.1\\
73P/SW3-B & JFC & 0.54 & 0.29 & $<$ 0.09 & 0.14 & 0.03 & 0.17 & ($<$
4.1) & ($<$ 19) & \nodata\\
73P/SW3-C & JFC & (0.49) & (0.22)  & $(<$ 0.16) & (0.12) & (0.03) &
(0.11) & $<$ 0.25 & 0.53 & \nodata\\
81P/Wild 2 & JFC & 0.9 & 0.27 & 0.6 & 0.22 & 0.15 & 0.45 & \nodata & 
\nodata & \nodata\\
103P/Hartley 2 & JFC & 1.95 & 0.24 & 0.66 & 0.13 & 0.10 & 0.75 & $<$
0.47 & 0.3 & \nodata\\
8P/Tuttle & OCC & 2.0 & 0.07 & \nodata & $<$ 0.04 & 0.04 & 0.26 & 0.37 & 0.4
& \nodata\\
153P/Ikeya-Zhang & OCC & 2.9 & 0.21 & \nodata & 0.83 & 0.21 & 0.57 &
0.5 & 5.7 & \nodata\\
C/1995 O1 (Hale-Bopp) & OCC & \nodata & 0.36 & \nodata & \nodata &
0.28 & 0.62 & 1.22 & 26.2 & 0.4\\
C 1996 B2 (Hyakutake) & OCC & \nodata & 0.19 & \nodata & \nodata & 0.20 & 0.61 &
0.95 & 18.2 & 0.2 \\
C/1999 H1 (Lee) & OCC & 1.9 & 0.22 & 0.7 & 0.7 & 0.25 & 0.63 & 1.22 &
1.6 & $<$ 3.6\\
C/1999 S4 (LINEAR) & OCC & $<$ 0.2 & 0.09 & \nodata & \nodata & $<$ 0.13 & 0.09
& 0.15 & 0.58 & $<$ 3.6 \\
C/1999 T1 (McNaught-Hartley) & OCC & 1.7 & 0.37 & \nodata & \nodata & \nodata & 0.65 &
1.4 & 17 & \nodata\\
C/2000 WM$_1$ (LINEAR) & OCC & 0.95 & 0.14 & \nodata & 0.2 & $<$ 0.05 & 0.47 &
0.35 & 0.48 & \nodata\\
C/2001 A2 (LINEAR) & OCC & 2.97 & 0.47 & \nodata & 0.15 & 0.37 & 1.6 & 1.48 &
3.9 & \nodata\\
C/2002 T7 (LINEAR) & OCC & 3.4 & \nodata & \nodata & 0.79 & \nodata & \nodata &
\nodata & 1.9 & 0.04 \\
C/2003 K4 (LINEAR) & OCC & 1.83 & 0.07 & $<$ 0.55 & $<$ 0.07 & $<$
0.04 & 0.41 & 0.86 & \nodata & \nodata \\
C/2004 Q2 (Machholz)& OCC  & 1.52 & 0.15 & 0.31 & 0.16 & 0.07 & 0.54 & 1.37
& 5.07 & \nodata\\
C/2006 M4 (SWAN) & OCC & 3.28 & \nodata & \nodata & \nodata & \nodata & 0.49
& 0.82 & 0.5 & \nodata\\
C/2006 P1 (McNaught)& OCC & \nodata & 0.24 & 1.5 & 0.49 & 0.45 & 0.47 & 0.42 &
1.8 & \nodata\\
C/2007 N1 (Lulin) & OCC & 3.72 & 0.14 & 0.24 & 0.12 & 0.07 & 0.68 &
1.19 & 2.18 & \nodata\\
C/2007 W1 (Boattini) & OCC & 3.69 & 0.5 & 1.74 & $<$ 0.12 & 0.29 & 1.97 &
1.57 & 4.50 & \nodata\\
C/2009 P1 (Garradd) & OCC & 2.74 & 0.25 & 0.48 & 0.09 & 0.07 & 0.82 &
0.95 & 8.9 & $<$ 0.2\\
C/2012 F6 (Lemmon)\tablenotemark{a}& OCC & 
1.48 & 0.19 & 0.52 & $<$ 0.12 & $<$ 0.05 & 0.29 & \nodata & \nodata &
\nodata 
\\
C/2012 F6 (Lemmon)\tablenotemark{b} & OCC &
\nodata & \nodata & \nodata & 0.54 & \nodata & \nodata & 0.67 & 4.03 & \nodata
\\
C/2012 S1 (ISON)\tablenotemark{c} & OCC &
1.13 & 0.07 & $<$ 0.95 & 0.16 & 0.11 & 0.27 & 0.32 & 1.37  & 0.16\\
C/2012 S1 (ISON)\tablenotemark{d} & OCC &
\nodata & 0.28 & 3.63 & 1.1 & 0.24 & \nodata & \nodata & \nodata & 
\nodata \\
C/2013 R1 (Lovejoy) & OCC & 2.29 &  0.25 & 0.1 & $<$0.06 & $<$ 0.07 &
0.59 & 0.92 & 11.3 & 0.034\\
\enddata
\tablerefs{$^a$Measurements at $R_b$ $>$ 1.2 au, 
$^b$Measurements at $R_b$ = 0.75~au,
$^c$Measurements at $R_b$ $>$ 0.83 au,
$^d$Measurements at $R_b$ $<$ 0.59 au}
\end{deluxetable}
\newpage

\section{\label{app:lippi}The new compilation of comet compositions from Lippi et al. (2021, 2020)}

While this paper was in review Lippi et al. (2021, 2020) (hereafter L20/21) published an important reanalysis of the composition of 20 comets, including all of those from the DR16 sample to which we compare our disk model outputs. The same eight molecules were considered as were included in the DR16 survey. L20/L21 applied fluorescence models to interpret cometary emission and telluric transmittance models accounting for attenuation in the Earth's atmosphere, which have been substantially improved since about 2010 \citep[][and references therein]{villa13}. The DR16 survey summarized work since the early stages of infrared high-resolution observation of comets in the 1990s, so we carefully assessed how the new L20/L21 results affect our analyses and conclusions. 

The main conclusions from our work (Conclusions 1 - 4 in Section 10) are based on ensemble properties of comets: average abundances and abundance ranges for individual molecules among the studied comets. These conclusions are not affected by the new survey because it results in differences between measurements in individual comets, but similar ranges and averages in abundances. L20/L21 report median values, very similar to the average abundances from DR16 for HCN, NH$_3$, C$_2$H$_2$, C$_2$H$_6$, and CH$_4$, and sufficiently close for CH$_3$OH, thereby not affecting Conclusions 1-4 from this paper. The larger difference in CO (5.2 +/- 1.3 from DR16 vs. abundance median of 2.66 in L21) is strongly dependent on sample size because the CO/H$_2$O relative abundance varies three orders of magnitude among comets.

The last conclusion (No. 5) in our work is based on comparison between disk model outputs and measurements in the individual comets listed Table 8: 103P/Hartley 2,  9p/Tempel 1, C/1999 H1 (Lee), C/2009 P1 (Garradd),  C/2013 R1 (Lovejoy), C/2007 N1 (Lulin), C/2012 S1 (ISON), and C/2004 Q2 (Machholz). We therefore applied $\chi^2$ test to the measurements from L20/L21 (Table~\ref{tab:lippi}.  These do not alter our conclusions that combining CO-rich and CO-poor material is required to reproduce comet abundances, and that dynamics (mixing hot and cold material) plays a key role in determining the chemistry. 

  \begin{deluxetable}{llllll}
    \tablecaption{\label{tab:lippi}Best fit parameters to the L20/L21 data assuming
      comets are made from material from two radii in the fiducial model, one inside ($R_1$) and one outside ($R_2$) the snowline. Shown are the combinations of $R_1$ and $R_2$ with the lowest $\chi^2$ which also match the observed abundances of each molecule to within a factor of 10.  The contribution of each radius is given as a percentage of the total calculated composition (so percentage of $R_1$ given is 100 x $f_{R1}$ where $f_{R1}$ is the fraction of $R_1$ in Equation~\ref{eq:frac}). The composition of all of our comet sample can be matched by such combined models.}
      \tablewidth{0pt}
      \tablehead{
 \colhead{Comet} & \colhead{time} & \colhead{$R_1$ (\%)} &
\colhead{$R_2$ (\%)} & \colhead{$\chi^2$} 
}
\startdata
9P/Tempel 1     &      1.6 &  4 (95\%) & 19 (5\%) & 0.92  \\
103P/Hartley 2      &      1.6 &   4 (90\%)  &  17 (10\%) &     0.76 \\ 
C/1999 H2 (Lee)          &      1.6 &  4 (90\%) & 18 (10\%)  &  2.36\\
C/2004 Q2 (Machholz)     &      2.0 &  4 (85\%)  & 20 (15\%) &  3.38   \\
C/2007 N1 (Lulin)        &      2.0 & 3 (85\%) &  19 (15\%)   & 2.20 \\
C/2009 P1 (Garradd)      &      2.0 & 3 (70\%)  & 20 (30\%) & 7.3 \\
C/2012 S1 (ISON)         &      2.0 & 4 (90\%) & 18 (10\%) & 3.57 \\
C/2013 R1 (Lovejoy)      &      2.0 & 3 (75\%) & 21 (25\%) & 6.2 \\
\enddata
\end{deluxetable}

The reason that our conclusions even on individual comets included in our study hold is plausibly explained by comparing DR16 and L20/L21 results. While significant differences are present for some molecules, most notably H$_2$CO, for each comet in Table~\ref{tab:lippi} abundances of four or more molecules (out of eight) either agree within reported uncertainties, or are sufficiently close, to not affect our conclusion 5. Comet C/2007 N1 has the best agreement - virtually all species, but even comets analyzed much earlier (C/1999 H1 and 9P) show similarities in at least half the measured abundances.

\newpage

\end{document}